\documentclass[10pt,letterpaper]{article}
\usepackage{arxiv}
\usepackage{amsmath,amssymb}
\usepackage{changepage}
\usepackage[utf8x]{inputenc} 

\usepackage{textcomp,marvosym}
\usepackage{cite}  
\usepackage[color=green!20]{todonotes}

\usepackage{nameref,hyperref}
\usepackage[right]{lineno}
\usepackage{microtype}
\DisableLigatures[f]{encoding = *, family = * }
\usepackage{color}

\definecolor{ForestGreen}{rgb}{0.1367, 0.5469, 0.1367}

\raggedright
\usepackage[aboveskip=1pt,labelfont=bf,labelsep=period,justification=raggedright,singlelinecheck=off]{caption}

\makeatletter
\renewcommand{\@biblabel}[1]{\quad#1.}
\makeatother
\usepackage{cleveref}

\usepackage{ulem}
\usepackage{lastpage,fancyhdr,graphicx}
\usepackage{subfig}
\usepackage{floatrow}   
\usepackage{epstopdf}
\usepackage{amsfonts}
\usepackage{amssymb}
\usepackage{amsmath}
\usepackage{mathrsfs}
\usepackage{subfig}
\usepackage{caption} 
\usepackage{graphicx}
\usepackage{psfrag} 
\usepackage{afterpage} 
\usepackage{nomencl}
\makenomenclature  
\fancyheadoffset[L]{2.25in}
\fancyfootoffset[L]{2.25in}
\def\bA{\mbox{\boldmath$ A$}}

\def\bN{\mbox{\boldmath$ N$}}

\def\bX{\mbox{\boldmath$ X$}}

\def\ba{\mbox{\boldmath$ a$}}

\def\bn{\mbox{\boldmath$ n$}}

\def\bt{\mbox{\boldmath$ t$}}
\def\bu{\mbox{\boldmath$ u$}}

\def\bx{\mbox{\boldmath$ x$}}

\DeclareGraphicsExtensions{.png,.eps} %%flip this order to go between high and low resolution image formats
\begin{document}
\vspace*{0.2in}
%
%Title must be 250 characters or less.
%
\begin{flushleft}
{\Large
%\textbf\newline{
%Isogeometric modeling of Kirchhoff-Love shell kinematics: a computational framework for modeling complex mechanical deformation pathways in biomembranes}\\
\textbf\newline{Biomembranes undergo complex, non-axisymmetric deformations governed by Kirchhoff-Love kinematics and revealed by a three dimensional computational framework}
}
%Alternate: High fidelity computations of Kirchhoff-Love shell kinematics reveal progressive growth of asymmetry during complex biomembrane deformation processes
%Alternate: High fidelity computations of K-L shell kinematics reveal increasing loss of symmetry for complex biomembrane deformations
%Please use "sentence case" for title and headings (capitalize only the first word in a title (or heading), the first word in a subtitle (or subheading), and any proper nouns).
 
%\newline
%\\
Debabrata Auddya\textsuperscript{1,*},
Xiaoxuan Zhang\textsuperscript{2,*},
Rahul Gulati\textsuperscript{1},
Ritvik Vasan\textsuperscript{3},
Krishna Garikipati\textsuperscript{2,4,5},
Padmini Rangamani\textsuperscript{3},
Shiva Rudraraju\textsuperscript{1**}
\\
\bigskip
\textbf{1} Department of Mechanical Engineering, University of Wisconsin-Madison, Madison, WI 53706, USA
\\
\textbf{2} Department of Mechanical Engineering, University of Michigan, Ann Arbor,
MI 48109, USA  
\\
\textbf{3} Department of Mechanical and Aerospace Engineering, University of California San Diego, La Jolla CA 92093, USA
\\
\textbf{4} Department of Mathematics, University of Michigan, Ann Arbor,
MI 48109, USA
\\
\textbf{5} Michigan Institute for Computational Discovery \& Engineering, University of Michigan, Ann Arbor
\bigskip

% Insert additional author notes using the symbols described below. Insert symbol callouts after author names as necessary.
% 
% Remove or comment out the author notes below if they aren't used.
%
% Primary Equal Contribution Note

% Additional Equal Contribution Note
% Also use this double-dagger symbol for special authorship notes, such as senior authorship.
%\ddag These authors also contributed equally to this work.

%\textcurrency Current Address: Dept/Program/Center, Institution Name, City, State, Country % change symbol to "\textcurrency a" if more than one current address note
% \textcurrency b Insert second current address 
% \textcurrency c Insert third current address


% Group/Consortium Author Note

% Use the asterisk to denote corresponding authorship and provide email address in note below.
* Both these authors contributed equally\\
**Corresponding Author \\
Email: shiva.rudraraju@wisc.edu
\end{flushleft}
\section*{Abstract}
Biomembranes play a central role in various phenomena like locomotion of cells, cell-cell interactions, packaging and transport of nutrients, transmission of nerve impulses, and in maintaining organelle morphology and functionality. 
During these processes, the membranes undergo significant morphological changes through deformation, scission, and fusion. 
Modeling the underlying mechanics of such morphological changes has traditionally relied on reduced order axisymmetric representations of membrane geometry and deformation. 
Axisymmetric representations, while robust and extensively deployed, suffer from their inability to model symmetry breaking deformations and structural bifurcations. 
To address this limitation, a three-dimensional computational mechanics framework for high fidelity modeling of biomembrane deformation is presented.
The proposed framework brings together Kirchhoff-Love thin-shell kinematics, Helfrich-energy based mechanics, and state-of-the-art numerical techniques for modeling deformation of surface geometries. 
Lipid bilayers are represented as spline-based surface discretizations immersed in a three-dimensional space; this enables modeling of a wide spectrum of membrane geometries, boundary conditions, and deformations that are physically admissible in a 3D space. 
The mathematical basis of the framework and its numerical machinery are presented, and their utility is demonstrated by modeling three classical, yet non-trivial, membrane deformation problems: formation of tubular shapes and their lateral constriction, Piezo1-induced membrane footprint generation and gating response, and the budding of membranes by protein coats during endocytosis. 
For each problem, the full three dimensional membrane deformation is captured, potential symmetry-breaking deformation paths identified, and various case studies of boundary and load conditions are presented. 
Using the endocytic vesicle budding as a case study, we also present a ``phase diagram'' for its symmetric and broken-symmetry states.

% Please keep the Author Summary between 150 and 200 words
% Use first person. PLOS ONE authors please skip this step. 
% Author Summary not valid for PLOS ONE submissions.   
%\section*{Author summary}
%\linenumbers

% Use "Eq" instead of "Equation" for equation citations.
\section*{Introduction}
\begin{figure}[!htb]
\centering 
  \psfrag{a}{(a) Filopodial protrusion}
  \psfrag{x}{\small $\text{u}_\text{x}=0$}
  \psfrag{z}{\small $\text{u}_\text{z}=0$}
  \psfrag{p}{\small $\phi=0$}
  \psfrag{y}{\small $\text{u}_\text{y}=0$}
  \psfrag{l}{\small $\gamma > 0$}
  \psfrag{c}{(c) Budding during endocytosis}
  \psfrag{b}{\shortstack{(b) Piezo1-induced membrane \\footprint generation}}
  \psfrag{f}{\small $\text{F}_\text{y}=0$}
  \psfrag{n}{\small Nucleus}
  \psfrag{v}{\small Vacuole}
  \psfrag{e}{\small \shortstack{Endoplasmic \\reticulum}}
  \psfrag{g}{\small Golgi}
  \includegraphics[width=0.8\textwidth]{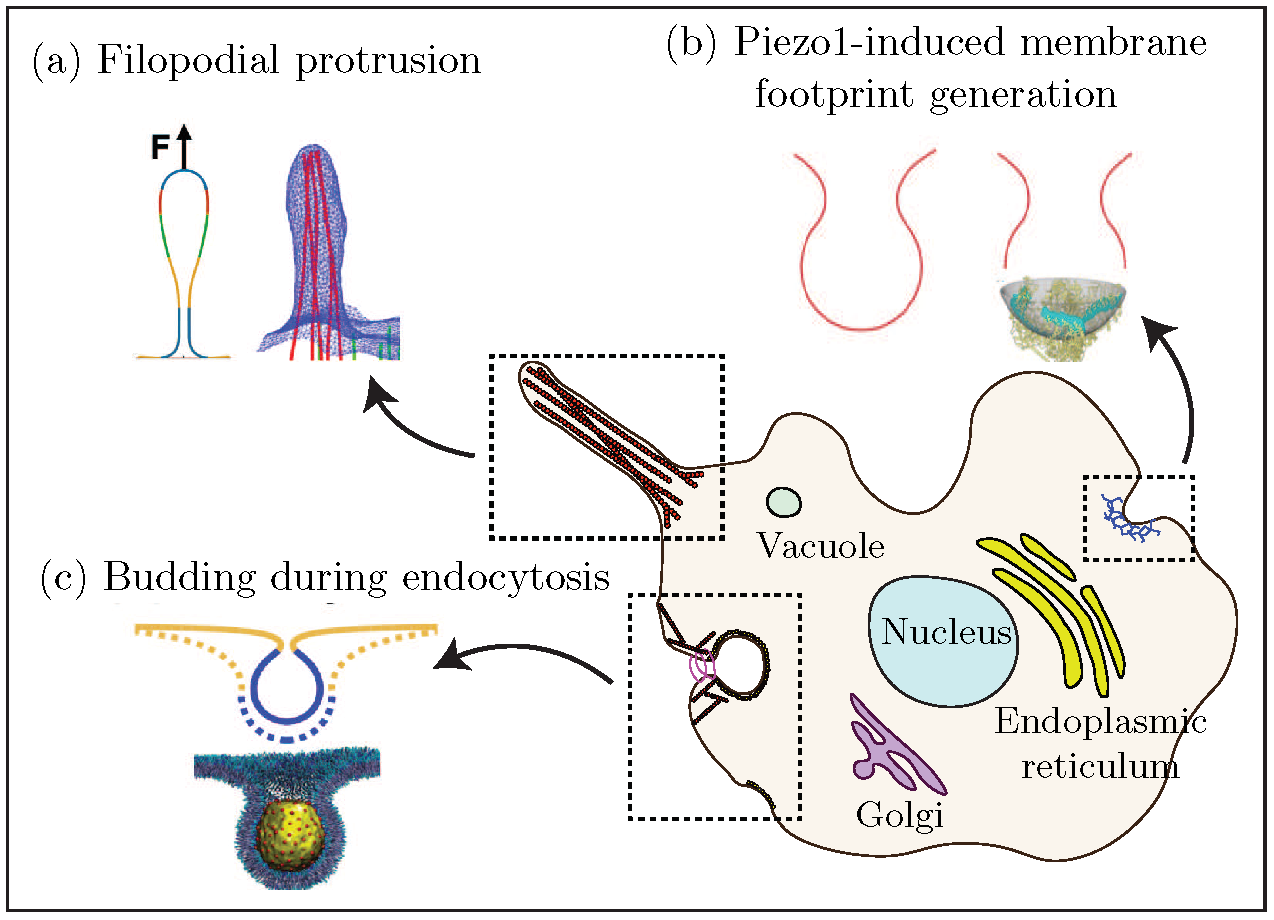}
\caption{Schematic of the various membrane biophysical phenomena modeled in this work to demonstrate the computational framework: (a) membrane tube pulling during filopodial protrusion, (b) dome formation and membrane footprint generation due to Piezo1 interaction, and (c) spontaneous curvature driven bud formation during endocytosis. Shown in insets are the schematic of the membrane deformation induced by the underlying protein complexes and its line diagram representation. } 
\label{fig:schematic_1}
\end{figure}

Membrane curvature is ubiquitous in biology \cite{mcmahon2005membrane}. 
Indeed, the bending of cell membranes is a central aspect of function for cells and organelles in many cellular processes such as cell migration \cite{zhao2013exo70}, cell membrane repair \cite{boye2017annexin}, membrane trafficking \cite{liu2009mechanochemistry} and cytokinesis \cite{schroeder1972contractile}, as well as the maintenance of distinctive membrane shapes within internal organelles like the endoplasmic reticulum \cite{hu2008membrane, shibata2006rough} and the Golgi complex \cite{mcniven2006vesicle}. 
Some important curved structures include tubules, sheets, vesicles and cisternae \cite{voeltz2007sheets}. 
A number of mechanisms have been identified to influence membrane bending, including geometric confinement by protein or lipid components of the membrane (intrinsic factors) \cite{zimmerberg2006proteins, koster2003membrane} and peripheral proteins and the cytoskeleton (extrinsic factors) \cite{takano2008efc, cocucci2012first}.
These mechanisms are often coupled and are spatio-temporally regulated by biochemical signaling cascades, leading to the mechanochemical coupling of signaling and membrane deformations. 
Lipid bilayer models that assume an in-plane fluid-like behaviour and an out-of-plane solid-like behaviour have provided notable insight to investigations of such curvature generation mechanisms. 
Particularly, the Helfrich-Canham model \cite{helfrich1973elastic} has furnished mechanistic insight to shape formation of liquid shells during vesiculation \cite{miao1991equilibrium, hurley2010membrane}, tubulation \cite{derenyi2002formation}, viral budding \cite{tzlil2004statistical}, clathrin-mediated endocytosis \cite{hassinger2017design}, and membrane neck formation \cite{alimohamadi2018role, Vasan2019}.
These modeling efforts have been complementary to advances in imaging techniques \cite{choi2009three, dupire2012full,Kukulski2012}, enabling a deeper appreciation of the complexity of membrane deformation.
\\
Despite the wealth of information provided by theoretical membrane mechanics models, an important restriction in several of these studies is the assumption of various degrees of symmetry for the membrane geometry and its deformation.
Indeed, the computation of membrane bending phenomena is significantly simplified with the axisymmetric assumption, but as we have shown recently \cite{Vasan2019}, this may come at the cost of generality and precision in identifying the underlying physics, as lower-energy, low-symmetry kinematic modes and even entire mechanisms may be overlooked.  
With growing interest in curvature-mediated biophysical phenomena and in 3D imaging and reconstruction methods \cite{Lee2020-kq,Lee2020-sb}, there is a need for general purpose computational tools to enable fully three dimensional numerical simulations.
\\
The continuum mechanical treatment of solids considers deformation as a mapping of the geometry (3D volume, 2D surface, or 1D curve) from its reference, undeformed configuration to a deformed current configuration under the influence of internal or external loads, of which the latter also may appear as boundary conditions. 
In limited cases, the geometry, loads and boundary conditions result in a mathematical problem of deformation of a $k$-manifold immersed in a $n$-dimensional space ($\mathbb{R}^n$). 
A 3-manifold is a volume, 2-manifold is a surface and 1-manifold is a curve. 
For $k=n$, modeling solid deformation is relatively straightforward and can be accomplished in the framework of Euclidean geometry using a rectilinear coordinate basis.
However, deformation of shell-like surface geometries, as is the case with biological membranes, involves tracking the underlying kinematics and evolution of geometric configurations of a $2$-manifold embedded in a $3$-dimensional space \cite{novozilov1959}. Such a geometric embedding  demands a non-Euclidean framework with a curvilinear coordinate basis. 
While the mathematical treatment of such a framework is well-developed (beginning with the celebrated work on differential geometry by Riemann in the $19^{th}$ century \cite{riemann1854hypotheses}), its application to three-dimensional modeling of biomembranes, which entails solving nonlinear partial differential equations in a curvilinear coordinate basis is relatively recent. Beginning with finite element models of Mindlin–Reissner plates \cite{hrabok1984review, arnold2005family, shi1991efficient, zienkiewicz1988robust} and Kirchhoff-Love shells \cite{simo1989stress, hrabok1984review, yang2000survey, bacsar1990finite}, initial efforts focused on developing numerical models in a rectilinear coordinate basis with approximated geometries and kinematics. 
However, the advent of spline-based geometric representations of surfaces and the more recent development of Isogeometric Analysis (IGA) techniques \cite{Cottrell2009} allow for an exact representation of surface geometries and the use of a curvilinear coordinate basis. 
Such treatments are now gaining traction in modeling structural applications \cite{kiendl2009isogeometric, kiendl2015isogeometric, zareh2019kirchhoff, nguyen2015extended} and also in the context of biological materials \cite{tepole2015isogeometric, sauer2017, roohbakhshan2017efficient, Duong2017}. 
We build upon these developments, especially from Sauer et al. \cite{sauer2017}, by adopting spline-based representations of surface geometries, treatments of membrane kinematics using a curvilinear basis, and the framework of IGA to develop a comprehensive computational modeling framework for studying complex deformations in biological membranes.\\
In this work, we present a three-dimensional, Helfrich-energy based, Kirchhoff-Love thin-shell computational framework for modeling the deformation of biological membranes in the regime of fully nonlinear kinematics and precise geometric representations. 
With this treatment, we are able to model membrane deformations, resolve geometric bifurcations, and explore post-bifurcation responses.
The main ingredients of this framework are the classical Kirchhoff-Love membrane kinematics \cite{novozilov1959}, a variational formulation of the governing equations underlying Helfrich-energy based mechanics on surface manifolds \cite{Duong2017, sauer2017,steigmann1999fluid}, and the numerical framework of IGA for solving the underlying partial differential equations. 
IGA methods form a numerical framework for finding approximate solutions to general partial differential equations \cite{Cottrell2009}, are a generalization of the classical Finite Element Method \cite{ciarlet2002finite, brenner2007mathematical, strang1973analysis}, and posses good numerical approximation and stability properties \cite{bazilevs2006isogeometric}. 
Crucially for accurate modeling of membrane biophysics, since IGA uses spline basis functions to represent the geometry and its deformation, it admits the continuity of slopes that is a characteristic of membranes in all states except for those of actual scission. 
As a result, we can now investigate simulations of membrane deformation under conditions that are notably more general (having fewer restrictive kinematic assumptions) than those considered previously in the literature \cite{alimohamadi2018role, bovzivc2001shapes, guckenberger2017theory, zheng1993helfrich, jian1993shape, molina2020diffusion}. 
The computational framework is implemented as an open-source software library and provided as a resource to the biophysics community \cite{GitRepo2020}.\\
To demonstrate the scope of the computational framework, we simulate three classical and non trivial membrane deformation phenomena  (see \Cref{fig:schematic_1}): (a) formation of tubular shapes and their lateral constriction, (b) Piezo1-induced membrane footprint generation and gating response, and (c) the budding of membranes due to the spontaneous curvature of the protein coats during endocytosis.
For each case, three dimensional membrane deformation is tracked, symmetry-breaking deformation pathways identified, and a few case studies of boundary conditions and loading are presented to exhibit the fidelity and modeling potential of the proposed methodology. 
We also present a phase diagram of symmetric and broken-symmetry states of membrane budding during endocytosis.\\
In the following sections, we present an outline of the mathematical framework and the model development, followed by a presentation of the three boundary values problems considered, their modeling results and biophysical implications. Finally, a discussion of the framework and its utility is presented in the conclusion. 

%%%%%%%%%%%%%%%%%%%%%%%%%%%%%%%%%%%%%%
%%%%%%%%%%%%%%%%%%%%%%%%%%%%%%%%%%%%%%
%BVP Schematics

%%%%%%%%%%%%%%%%%%%%%%%%%%%%%%%%%%%%%%%%%%%%%%%%%%%%%%%%%%%%%
\begin{figure}[!p]
\centering
  \psfrag{x}{\small $y$}
  \psfrag{y}{\small $x$}
  \psfrag{z}{\small $z$}
  \psfrag{a}{(a) Membrane tube pulling and tube constriction}
  \psfrag{b}{(b) Piezo1-induced membrane footprint generation}
  \psfrag{c}{(c) Bud formation due to spontaneous curvature}
  \psfrag{d}{Tube pulling}
  \psfrag{e}{Tube constriction}
  \psfrag{f}{\small \shortstack[c]{constriction \\pressure}}
  \psfrag{g}{$\text{u}_\text{x}=0$}
  \psfrag{h}{$\text{u}_\text{y}$}
  \psfrag{u}{$\text{u}_\text{y}=0$}
  \psfrag{w}{$\phi=70$\textdegree}
  \psfrag{v}{$\text{t}_\text{y} < 0$}
  \psfrag{i}{$\text{u}_\text{z}=0$}
  \psfrag{j}{$\text{u}_\text{y}=0$}
  \psfrag{k}{$\phi=0$}
  \psfrag{l}{\small \shortstack[c]{membrane \\footprint}}
  \psfrag{m}{$\bt=\gamma \tilde \bn$}
  \psfrag{n}{$\text{t}_\text{y}>0$ }
  \psfrag{o}{$\text{t}_\text{y} \approx 0$}
  \psfrag{p}{$\text{t}_\text{y}$}
  \psfrag{r}{\small \shortstack[c]{coat area \\$h_0>0$}}
  \psfrag{s}{$\phi$}
  \includegraphics[width=0.93\textwidth]{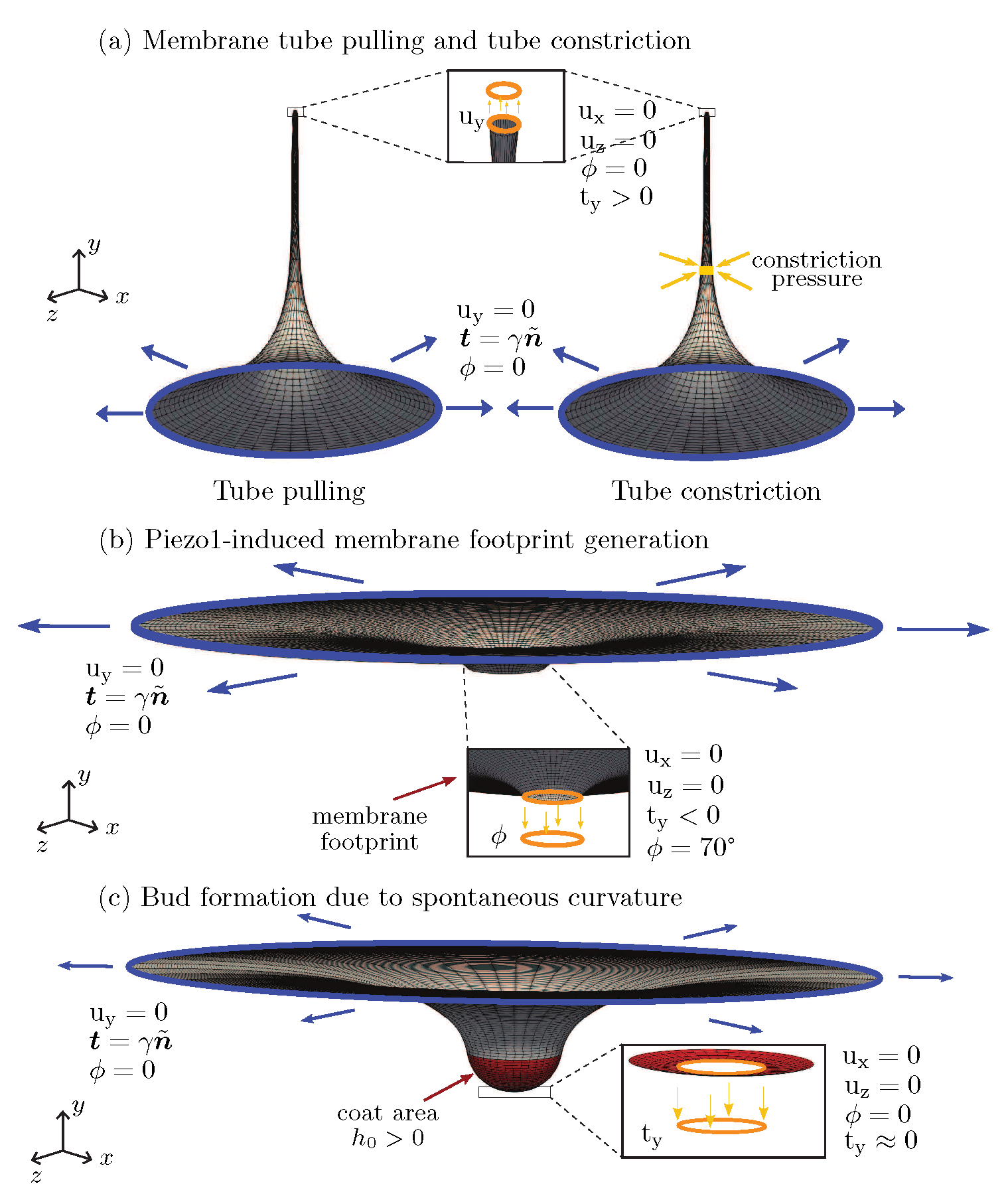}
\caption{Schematic of the various membrane boundary value problems considered in this work. Shown are the geometry and boundary conditions for (a) formation of tubular shapes and their lateral constriction due to the application of axisymmetric constriction pressure, (b) Piezo1-induced membrane footprint generation, and (c) the budding of membranes due to the spontaneous curvature of the protein coats during endocytosis. Here, $u_x$, $u_y$ and $u_z$ are the displacement components, $\bt$ is the surface traction and $t_y$ its component along the $y$-axis, $\tilde \bn$ is the normal to the boundary curve, $\phi$ is the boundary slope, $\gamma$ is the surface tension applied on the membrane boundary, and $h_0$ is the instantaneous curvature. Blue and orange colors identify the outer and inner rims, respectively.}
\label{fig:schematict}
\end{figure}
%%%%%%%%%%%%%%%%%%%%%%%%%%%%%%%%%%%%%%
%%%%%%%%%%%%%%%%%%%%%%%%%%%%%%%%%%%%%%
\section*{Materials and Methods}
The mathematical framework consists of surface geometry parametrization, Kirchhoff-Love membrane kinematics, Helfrich-energy based mechanics of lipid bilayers and surface partial differential equations governing mechanical deformation. 
Key ingredients of this framework are described in the mathematical framework subsection below, while the detailed mathematical derivations are provided in the SI. Using the IGA apparatus, the mathematical treatment is then cast into a numerical formulation that allows for solving the governing equations to obtain the spatial evolution of membrane deformation. 
These aspects of the framework are discussed under the computational implementation subsection. 

\subsection*{Mathematical framework}
The mathematical treatment introduced here follows from Sauer et al. \cite{sauer2017}. Only the important results are summarized in this section, and the detailed derivations are presented in the SI. 
\subsubsection*{Surface parametrization and kinematics}
\begin{figure}[!htb]
 \centering
  \psfrag{a}{$\bA_1$}
  \psfrag{b}{$\bA_2$}
  \psfrag{c}{$\bN$}
  \psfrag{d}{$\Omega_0$}
  \psfrag{e}{$\partial \Omega_0$}
  \psfrag{f}{$\bA_1 = \frac{\displaystyle \partial \bX}{\displaystyle \partial \xi^1}$}
  \psfrag{g}{$\bA_2 = \frac{\displaystyle \partial \bX}{\displaystyle \partial \xi^2}$}
  \psfrag{h}{$\bN = \frac{\displaystyle \bA_1 \times \bA_2}{\displaystyle \left\lVert \bA_1 \times \bA_2 \right\rVert}$}
  \psfrag{i}{$\bX = \bX(\xi^1,\xi^2) \in \Omega_0 $}
  \psfrag{m}{$\ba_1$}
  \psfrag{n}{$\ba_2$}
  \psfrag{o}{$\bn$}
  \psfrag{p}{$\Omega$}
  \psfrag{q}{$\partial \Omega$}
  \psfrag{r}{$\ba_1 = \frac{\displaystyle \partial \bx}{\displaystyle \partial \xi^1}$}
  \psfrag{s}{$\ba_2 = \frac{\displaystyle \partial \bx}{\displaystyle \partial \xi^2}$}
  \psfrag{t}{$\bn = \frac{\displaystyle \ba_1 \times \ba_2}{\displaystyle \left\lVert\ba_1 \times \ba_2 \right\rVert}$}
  \psfrag{u}{$\bx = \bx(\xi^1,\xi^2) \in \Omega $}
  \psfrag{z}{membrane deformation}
  \psfrag{j}{Reference configuration}
  \psfrag{v}{Current configuration}
  \includegraphics[width=\textwidth]{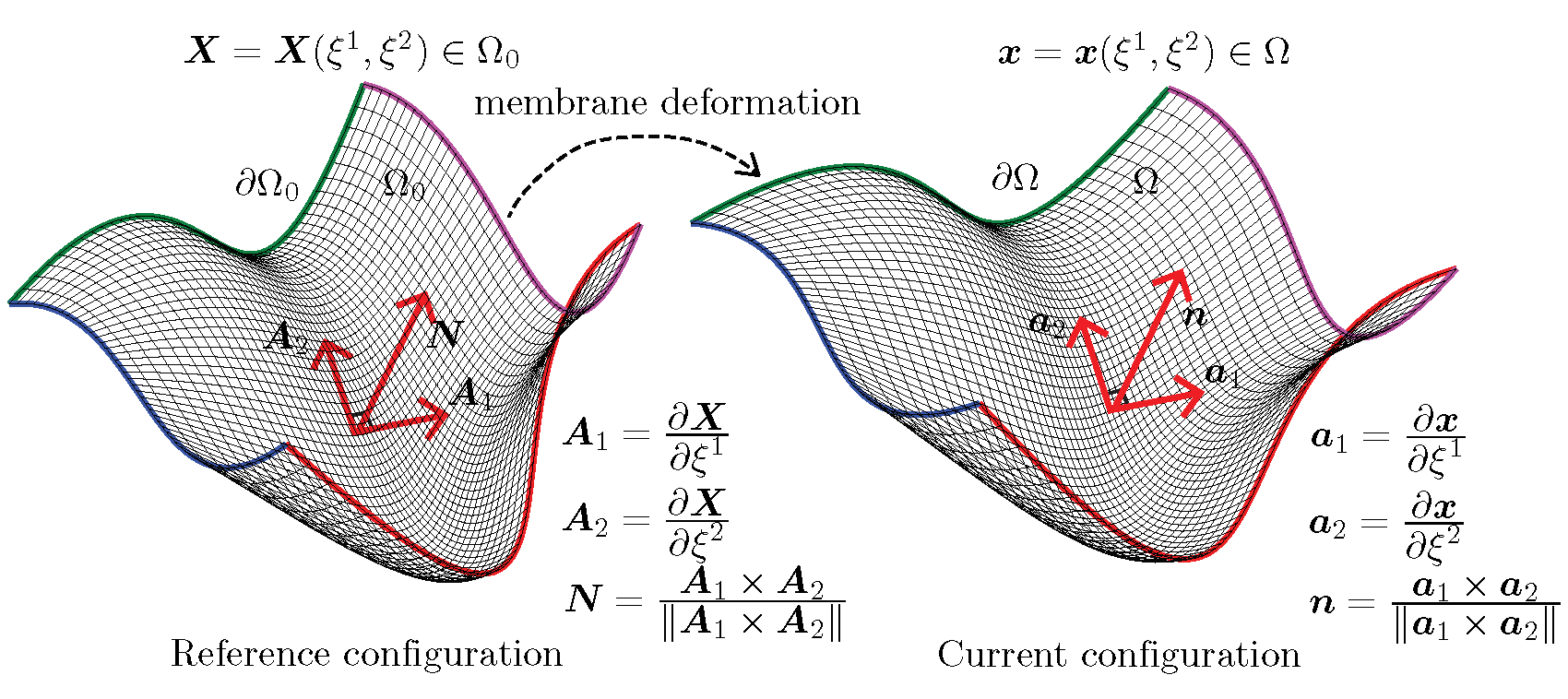}
\caption{Surface parametrization of a biomembrane in the reference undeformed configuration ($\Omega_0$) and current deformed configuration ($\Omega$). The 2D surface, $\Omega_0$, is bounded by the curves $\partial \Omega_0$ (highlighted with color), and embedded in a 3D volume. Here, $\bX$ is the position vector of a point on the surface parametrized in terms of the surface coordinates ($\xi^1$,$\xi^2$) which are associated with a flat 2D domain that is then mapped to $\Omega_0$ as $\bX=\bX(\xi^1,\xi^2)$. The local tangent vectors to the surface at $\bX$ are $\bA_1$ and $\bA_2$,  and $\bN$ is the corresponding surface normal. The position dependent triads \{$\bA_1$,$\bA_2$,$\bN$\} and \{$\ba_1$,$\ba_2$,$\bn$\} form the local curvilinear coordinate basis for the reference undeformed configuration and current deformed configuration, respectively. }
\label{fig:parametrization}
\end{figure}

Consider a lipid bilayer represented as a surface (2-manifold) embedded in a 3D volume, as shown in Fig. \ref{fig:parametrization}. 
Let the reference (undeformed) configuration and the current (deformed) configuration of the surface geometry be denoted by $\Omega_0$ and $\Omega$, respectively. The configurations $\Omega_0$ and $\Omega$ are parametrized by the coordinates $\xi^1$ and $\xi^2$ that map a flat 2D domain to the surface coordinates $\bX$ and $\bx$: 
\begin{equation}
\begin{aligned}[c]
\bX = \bX(\xi^1,\xi^2) \quad \forall  \quad \bX \in \Omega_0,
\end{aligned}
\qquad
\begin{aligned}[c]
\bx = \bx(\xi^1,\xi^2)  \quad \forall  \quad \bx \in \Omega.
\end{aligned}
\end{equation}

The (covariant) tangent vectors in the reference and current configuration are given by:
 \begin{equation}
\begin{aligned}[c]
\bA_{I} = \frac{\partial \bX}{\partial \xi^I}=\bX,_{I},
\end{aligned}
\qquad
\begin{aligned}[c]
\ba_{i}= \frac{\partial \bx}{\partial \xi^i} = \bx,_{i}.
\end{aligned}
\end{equation}
In the expressions that follow, except when indicated otherwise, uppercase letters are associated with the reference configuration and lowercase letters are associated with the current configuration. \\
Using the tangent vectors we define the surface normals as follows:
\begin{equation}
\begin{aligned}[c]
\bN=\frac{\bA_1 \times \bA_2}{\left\lVert \bA_1 \times \bA_2 \right\rVert},
\end{aligned}
\qquad
\begin{aligned}[c]
\bn=\frac{\ba_1 \times \ba_2}{\left\lVert \ba_1 \times \ba_2 \right\rVert}.
\end{aligned}
\end{equation}

From the triad consisting of the tangent vectors and the normal that form the local curvilinear coordinate basis, we can obtain expressions for the metric tensor,
\begin{equation}
\begin{aligned}[c]
A_{IJ}=\bA_{I} \cdot \bA_{J}
\end{aligned}
\qquad
\begin{aligned}[c]
a_{ij}=\ba_{i} \cdot \ba_{j} 
\end{aligned}
\end{equation}
The second order derivatives of the surface coordinates $\bX$ and $\bx$ are given by:
\begin{equation}
\begin{aligned}[c]
\bA_{I,J}=\frac{\partial \bA_I}{\partial \xi_J}
\end{aligned}
\qquad
\begin{aligned}[c]
\ba_{i,j}=\frac{\partial \ba_i}{\partial \xi_j}
\end{aligned}
\end{equation}
and from them we obtain the components of the curvature tensor,
\begin{equation}
\begin{aligned}[c]
B_{IJ} = \bA_{I,J} \cdot \bf{{N}}
\end{aligned}
\qquad
\begin{aligned}[c]
b_{ij} = \ba_{i,j} \cdot \bn
\end{aligned}
\end{equation}

%\needref. 
We are now able to define the primary kinematic metrics of interest: the mean and Gaussian curvature. The mean curvature and Gaussian curvature are frame invariant measures of a surface geometry, and hence are natural choice for representing the kinematics of the surface as it deforms. Using the components of the curvature tensor, we can obtain expressions for the mean curvature,
%\todokg[inline]{Cannot show this without defining the metric tensor and curvature tensor. And, we need to see the SI, also. It had all the details.}
\begin{equation}
\begin{aligned}[c]
H = \frac{1}{2} B^{IJ} A_{IJ}  \quad \text {on} \quad \Omega_0,
\end{aligned}
\qquad
\begin{aligned}[c]
h = \frac{1}{2} b^{ij} a_{ij}  \quad \text {on} \quad \Omega
\end{aligned}
\end{equation}
and the Gaussian curvature, 
\begin{equation}
\begin{aligned}[c]
K = \frac{\left|B\right|}{\left|A\right|}  \quad \text {on} \quad \Omega_0,
\end{aligned}
\qquad
\begin{aligned}[c]
\kappa = \frac{\left|b\right|}{\left|a\right|}  \quad \text {on} \quad \Omega, \quad \left| \cdot \right| = det(\cdot)
\end{aligned}
\end{equation}

%Parametrization

\subsubsection*{Biophysics of membrane deformation}
With a focus on representing the correct deformation, a biomembrane is often modeled as a thin elastic shell governed by the classical Helfrich formulation \cite{helfrich1973elastic, kishimoto2011determinants,morlot2012membrane} of membrane bending energy. In this treatment, the primary kinematic variables are the curvatures capturing the bending of the membrane, and the elastic energy density of the membrane is given by:
\begin{equation}
w = k_B (h-h_0)^2+k_G \kappa
\end{equation}
where $k_B$ and $k_G$ are the bending modulus and the Gaussian curvature modulus of the membrane, and $h_0$ represents the instantaneous curvature induced in the membrane.  \\
Furthermore, we assume that the membrane is area preserving (\textit{i.e} the membrane area is constant) \cite{evans1979mechanics} -- a constraint that is implemented using a Lagrange multiplier field. Enforcing the area-preserving condition results in the following expression for the elastic energy density:
\begin{equation}
w = k_B (h-h_0)^2+k_G \kappa + \lambda (J-1)
\end{equation}
where $\lambda$ is the point value of the Lagrange multiplier field, and $J$ is the surface Jacobian (ratio of area in the current configuration to the area in the reference configuration). Thus, while the Helfrich energy is defined entirely in terms of the geometry of the surface, the Lagrange multiplier, often interpreted as the membrane tension \cite{rangamani2014protein,steigmann1999fluid}; i.e. a non-geometric quantity, also plays a role in determining the minimum energy configuration. We ignore any fluid \cite{rangamani2013interaction, arroyo2009relaxation} and friction \cite{simunovic2017friction, quemeneur2014shape, rahimi2012shape} properties of the bilayer, guided by the dominance of unstable and stable equilibrium states over relaxation or rate processes. 
The augmented Helfrich Hamiltonian whose extremum is sought over the membrane surface, including the Lagrange multiplier field $\lambda$ is given as:
\begin{equation}
E = \int_{\Omega} (k_B (h-h_0)^2+k_G \kappa + \lambda (J-1))~da
\end{equation}
where $\Omega$ is the domain of integration over the membrane surface.

\subsubsection*{Governing equations}
The governing equation for quasi-static mechanical equilibrium in 3D simulations is obtained as the Euler-Lagrange condition of the Helfrich energy functional following standard variational arguments, and is given by: 
\begin{equation}
\int_{\Omega} \frac{1}{2} \delta a_{ij} \sigma^{ij} ~da + \int_{\Omega} \delta b_{ij} M^{ij} ~da - \int_{\Omega} \delta \boldsymbol{x} \cdot \boldsymbol{p} ~da - \int_{\partial \Omega} \delta \boldsymbol{x} \cdot \boldsymbol{t} ~ds = 0,
\label{EqnWeakForm}
\end{equation}
where $\partial \Omega$ is the membrane boundary on which surface tractions and displacement boundary conditions can be applied, as shown in \Cref{fig:parametrization}. Furthermore, $\delta a_{ij}$ and $\delta b_{ij}$ are variations of the components of the metric tensor and the curvature tensor, respectively, 

%\todokg[inline]{Do we need to point out the index lowering operation on $b_{ij}$?}

Here, $\sigma^{ij}$ are the components of the stress tensor, $M^{ij}$ are components of the moment tensor (in the current configuration), $\boldsymbol{p}$ is the pressure applied on the membrane surface (in the case of the tube constriction boundary value problem), and $\boldsymbol{t}$ is the surface traction. \\
For a hyperelastic material model, we can express the stress and moment components in terms of the strain energy density as \cite{Duong2017}: 
\begin{eqnarray}
\sigma^{ij} = \frac{2}{J} \frac{\partial w}{\partial a_{ij}},\\
M^{ij} = \frac{1}{J} \frac{\partial w}{\partial b_{ij}}
\end{eqnarray}
For the Helfrich type strain energy density, these take the form:
\begin{align}
\sigma^{ij} &= (k_B (h-h_0)^2 - k_G \kappa) a ^{ij} - 2 k_B (h-h_0) b^{ij},\\
M^{ij} &= (k_B (h-h_0) + 2 k_G h)a ^{ij} - k_G b^{ij}\
\end{align}

%%%%%%%%%%%%%%%%%%%%%%%%%%%%%%%%%
\subsection*{Computational implementation}
The governing equations of membrane deformation, given as a system of nonlinear elliptic partial differential equations, are solved using the framework of Isogeometric Analysis \cite{Cottrell2009}. 
An IGA method-based membrane mechanics framework has been developed for this work, and is built on top of the PetIGA \cite{2016petiga} open source library. In an IGA approach, the membrane geometry is discretized using a spline mesh and the governing equations are converted to a nonlinear system of equations, which is then solved to obtain the deformed membrane shape. 
Of importance to our central result is that this framework naturally admits both symmetric and asymmetric deformation modes driven by the underlying physics. 

We make three key important remarks about this framework.
First, a fundamental conjecture of the Helfrich model is that the characteristic length scales of the problem are much larger than the thickness of the bilayer \cite{helfrich1973elastic}. 
This assumption allows us to neglect the effect of transverse shear deformations and consider the classical Kirchhoff–Love shell kinematics for thin shell geometries \cite{novozilov1959}. 
Second, numerical solutions to the membrane shape equations in general coordinates are challenging because of continuity requirements in the numerical scheme. Specifically, the deformation must maintain first-order; i.e., $C^1$, spatial continuity of the membrane surface in all except extreme states, such as the  configuration of actual scission. %Numerical solutions to the governing equations obtained using Kirchhoff–Love shell theory are challenging due to the inherent requirement of a continuously differentiable ($C^1$-continuity) numerical basis. 
We overcome this challenge by adopting both spline basis functions, which allow high-order continuity, and the numerical framework of Isogeometric Analysis \cite{Cottrell2009}. 
Finally, an inherent limitation of the Helfrich energy formulation in three dimensional simulations is the lack of resistance to shear deformation modes.
The zero energy modes corresponding to shear deformation are eliminated in this framework by adding shear stabilization terms of smaller magnitude relative to the traditional bending terms in the Helfrich energy \cite{sauer2017}, thus restoring stability to the numerical model.

%%%%%%%%%%%%%%%%%%%%%%%%%%%%%%%%%
%%%%%%%%%%%%%%%%%%%%%%%%%%%%%%%%%
\section*{Results}
We demonstrate the simulation framework using three classical membrane deformation problems: formation of tubular shapes and their lateral constriction, Piezo1-induced membrane footprint generation and gating response, and the budding of membranes by protein coats during endocytosis. Through these examples, we also demonstrate the emergence of increasingly complex membrane deformations that are beyond the scope of traditional axisymmetric formulations. These problems are described in detail below. 
 
\subsection*{Formation of tubular shapes and their lateral constriction}
\begin{figure}[!p]
\centering
  \psfrag{x}{$x$}
  \psfrag{y}{$y$}
  \psfrag{z}{$z$}
  \psfrag{m}{\small $u_y$ (nm)}
  \psfrag{n}{\small $u_y$ (nm)}
  \psfrag{a}{(a)}
  \psfrag{b}{(b)}
  \psfrag{c}{(c)} 
  \psfrag{r}{\small $\gamma$} 
  \psfrag{s}{\small Axial force ($t_y$)}  
  \psfrag{u}{\small Tube profile with increasing axial force}  
  %\psfrag{v}{\small $r \propto \sqrt{\frac{\displaystyle 1}{\displaystyle \gamma}}$}  
  \psfrag{v}{} 
  \psfrag{d}{(d)}   
  \psfrag{e}{\small $\kappa_B=100 \text{pN}\cdot \text{nm}$}
  \psfrag{f}{\small $\gamma = 1.0 \text{pN}/ \text{nm}$}
  \psfrag{g}{\small $\kappa_B=100 \text{pN}\cdot \text{nm}$}
  \psfrag{h}{\small $\gamma = 0.75 \text{pN}/ \text{nm}$}
  \psfrag{i}{}
  \psfrag{j}{\small $\gamma = 1.0 \text{pN}/ \text{nm}$}
  \psfrag{k}{}
  \psfrag{l}{\small $\gamma = 2.0 \text{pN}/ \text{nm}$}
  \psfrag{o}{}
  \psfrag{p}{\small $\gamma = 3.0 \text{pN}/ \text{nm}$}
 \includegraphics[width=0.99\textwidth]{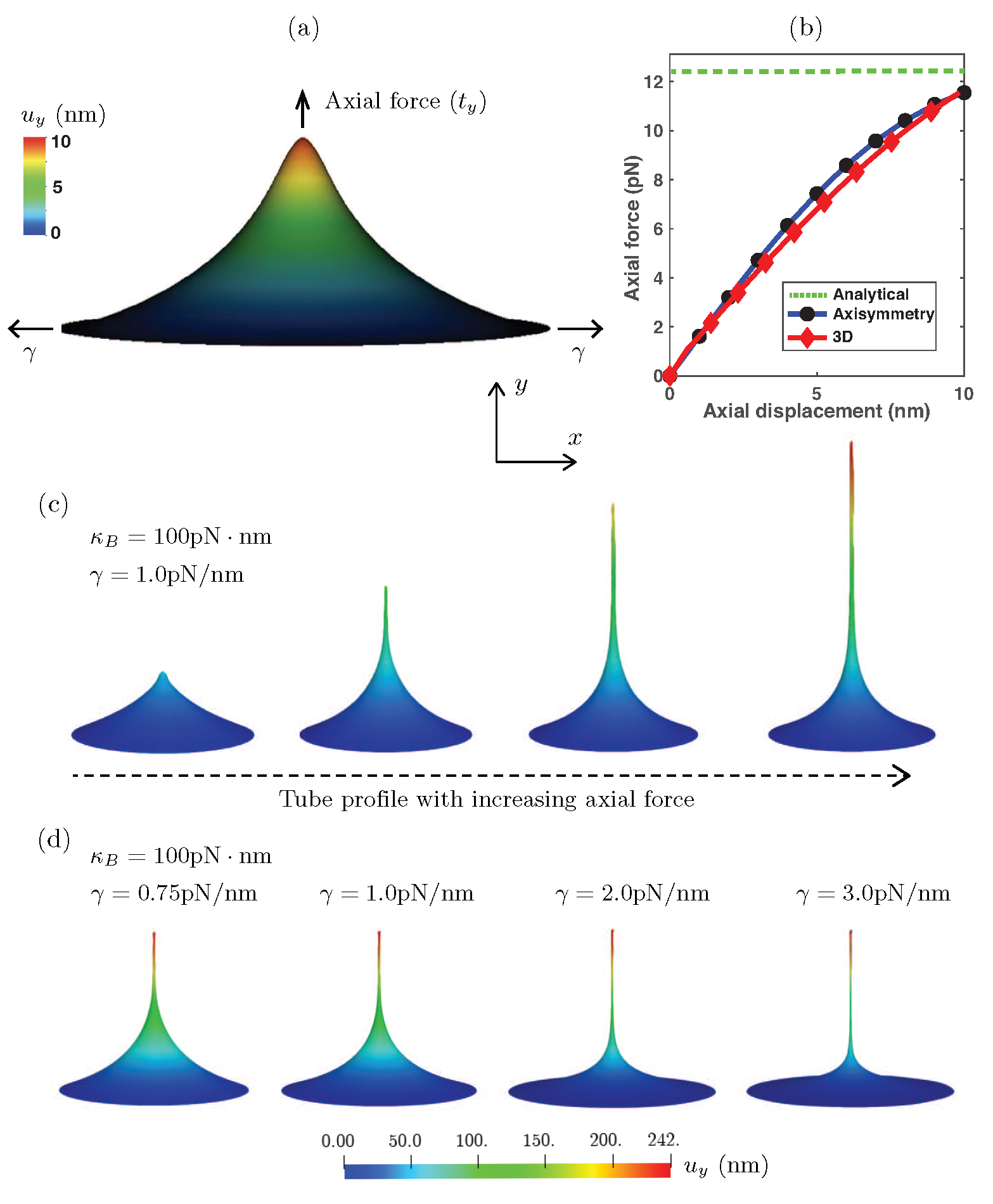}
\caption{Deformation profile and force-displacement response of a membrane during tube pulling. Shown are the (a) deformation profile with the application of axial force ($t_y$) on a membrane with a bending modulus ($\kappa_B$) of 20 pN-nm under a surface tension ($\gamma$) of 0.1 pN/nm, (b) comparison of the 3D force-dispalcment response with the axisymmetric solution and the equilibrium tube pulling force predicted by the analytical model, (c) progression of tube pulling with increasing axial force, and (d) dependence of the deformation profile and tube radius on the surface tension of the membrane.} 
\label{fig:pullout:BVP1}
\end{figure}

Many cell organelles and cytoplasmic projections are shaped as vesicles, tubes, or elongated membrane structures. 
Some examples of such shapes are the filopodia protrusions, inner mitochondrial region, endoplasmic reticulum, the Golgi complex, etc. (see ~\Cref{fig:schematic_1}). 
These tubular structures play an important role in the locomotion of cells, production and folding of proteins, and in the formation of vesicles for transporting proteins and lipids among others. 
A typical mechanism for producing these tubular shapes involves motor proteins that attach to the cell membrane and pull it along the filaments of the cytoskeleton \cite{Koster2003-xm, Shaklee2008-hf}. 
Further, as is the case with the fission of endocytic vesicles, the tubular or vesicular structures also undergo constriction by scission proteins like dynamin \cite{Simunovic2017-hj, Morlot2013-tk, Kong2018-rd}.
This constriction mediates a membrane pinch-off mechanism that leads to the formation of vesicles.
From a biophysical standpoint, it is important to gain a quantitative understanding of the interaction between the proteins and the membranes by determining the deformation mechanisms, forces exerted by proteins, and kinematic constraints.\\ 
A classic benchmark problem in the understanding of elongated biomembrane structures is the analytical model of the formation and interaction of membrane tubes proposed by Der\'enyi et al \cite{derenyi2002formation}. 
Some key results of this model are the prediction of the magnitude of protein-membrane interaction forces and tubule radius, and their dependence on the membrane bending modulus ($\kappa_B$) and surface tension ($\gamma$). 
The protein pulling force, $t_y$, and the tubule radius, $r$, are related to the bending modulus and surface tension of the membrane as follows: $t_y \propto \sqrt{\kappa_B~\gamma}$ and $r \propto \sqrt{\kappa_B/\gamma}$.
In addition to these analytical estimates, numerical solutions to the problem of membrane tube pulling, albeit with axisymmetric constraints on deformation, are available in the literature \cite{Lipowsky2012-zd, Bahrami2017-dg} and in our earlier work \cite{Vasan2019}.
%Given the paucity of analytical solutions to membrane deformation problems based on the Helfrich-energy model, 
We take advantage of the analytical estimates proposed by Der\'enyi et al., the numerical solutions available from axisymmetric models \cite{Vasan2019}, and validate the computational framework proposed in this work by comparing the load-displacement response of membrane tube pulling from these three approaches. \\
The boundary value problem solved, along with the spatial discretization (mesh), boundary conditions on the displacement ($\bu$) and the membrane boundary slope ($\phi$) are shown in ~\Cref{fig:schematict}(a).
The simulation results are shown in \Cref{fig:pullout:BVP1}: \Cref{fig:pullout:BVP1} (a) is the deformation profile obtained during tube pulling, and in \Cref{fig:pullout:BVP1}(b) is the load-displacement response of the 3D framework compared to the asymmetric result and the equilibrium value of tube pulling force predicted by the analytical model.
We note that the analytical model only predicts the final equilibrium value of the tube pulling force, and hence only a single value of the force from the analytical model is plotted. 
As can be seen from \Cref{fig:pullout:BVP1}(b), the 3D model very closely tracks the axisymmetric solution and asymptotically approaches the equilibrium value of force from the analytical solution.
Further, we show the evolution of the deformation profile with increasing tube pulling force in \Cref{fig:pullout:BVP1}(c), and the dependence of the deformation profile and tubule radius on the applied surface tension in \Cref{fig:pullout:BVP1}(d).\\

\begin{figure}[!htb]
\centering
  \psfrag{z}{\small \shortstack[c]{Tube drawn \\from a membrane\\ reservoir}}
  \psfrag{a}{\small \shortstack[c]{Tube geometry \\and the mesh}}
  \psfrag{b}{\small \shortstack[c]{Deformed shape}}
  \psfrag{c}{\small Progression of non-symmetric membrane constriction}
  \psfrag{d}{\small \shortstack[c]{Axisymmetric\\constriction\\pressure}}
  \psfrag{e}{\tiny \shortstack[c]{Mean \\curvature}}
 \includegraphics[width=\textwidth]{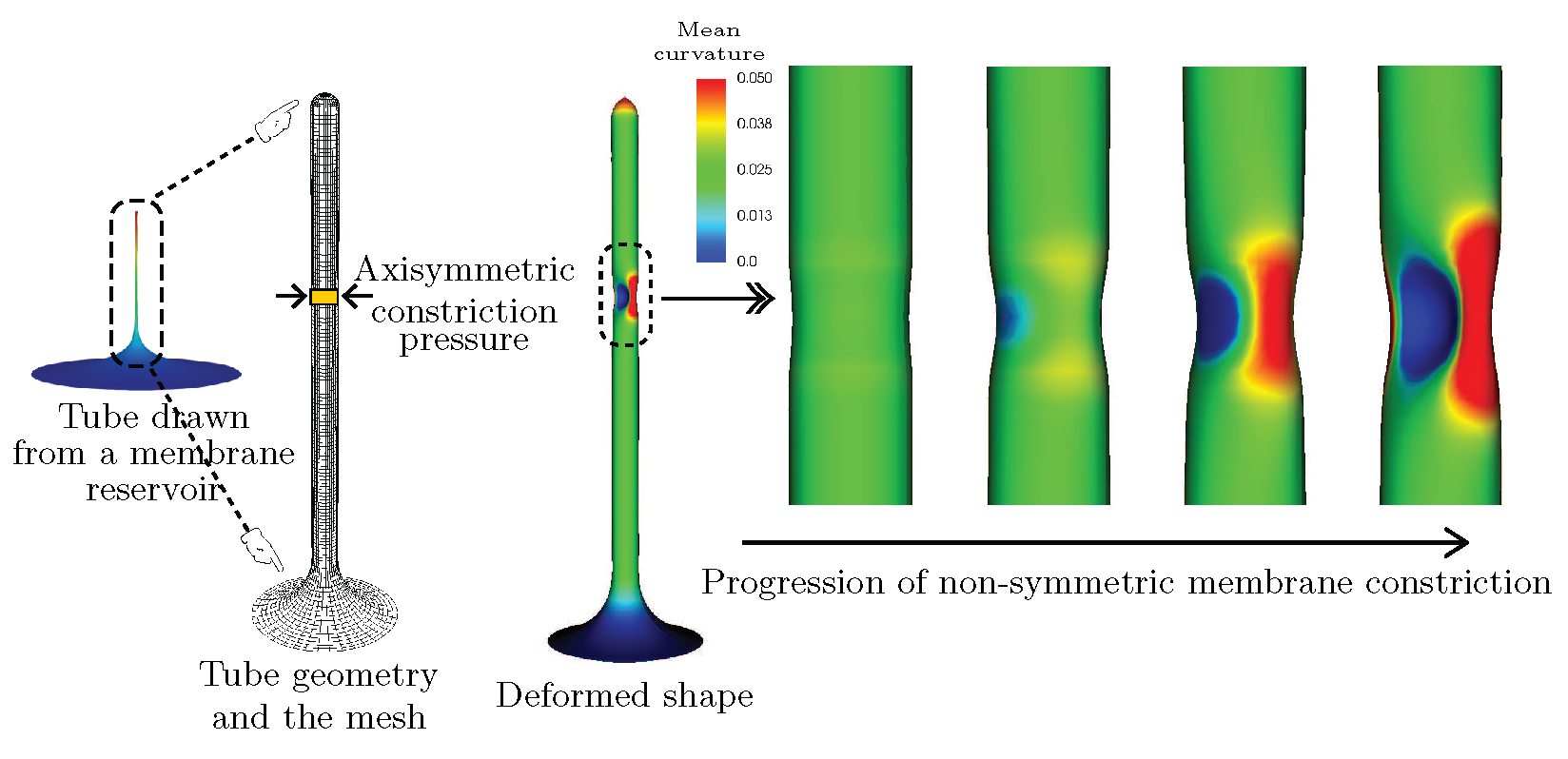}
\caption{Progression of membrane tube constriction with increasing constriction pressure leading to non-symmetric pinching profiles of deformation. Shown are the tube geometry and the computational mesh, and the progression of the non-symmetric membrane constriction due to the constriction pressure.}
\label{fig:pinching}
\end{figure}

We further consider the effect of lateral constriction pressure on the tubular geometry and demonstrate the non-axisymmetric pinching deformation profile that is predicted by the computational framework.
For this boundary value problem, we consider a tubular geometry (shown in ~\Cref{fig:schematic_1}(a), under tube constriction) and apply an axisymmetric constriction pressure that would be applied by a spiral collar protein like dynamin \cite{morlot2012membrane, roux2006gtp, danino2004rapid}. 
As can be expected, an axisymmetric model would predict an axisymmetric pinching profile in the vicinity of the constriction pressure \cite{Vasan2019}.
However, the fully 3D model considered in this computational framework is not limited to axisymmetric solutions, and is thus able to predict non-axisymmetric states when they are the energy minimizing solutions to the governing equations of membrane deformation. 
The progression of the non-axisymmetric solution with increasing constriction pressure is shown in ~\Cref{fig:pinching}.
This shape of the membrane has significant implication on the force and energy barrier of protein induced pinching of membranes, as has been studied in detail in our recent work demonstrating how non-axisymmetric buckling lowers the energy barrier associated with membrane neck constriction \cite{Vasan2019}. 
In that study, we used an earlier version of the computational framework proposed here to study the influence of location, symmetry constraints, and helical forces on membrane neck constriction in a lipid bilayer. 
Simulations from our model demonstrated that the energy barriers associated with constriction of a membrane neck are location-dependent, and if symmetry restrictions are relaxed, the energy barrier for constriction is dramatically lowered and the membrane buckles at lower values of the  constriction pressure. 
These studies helped establish that even though there exist different molecular mechanisms of neck formation in cells, the mechanics of constriction of a cylindrical membrane tubule naturally leads to a loss of symmetry that can lower the energy barrier to constriction.
This loss of symmetry may be a common mechanism for different scission processes and clearly demonstrates the need for fully three dimensional computational framework to give predictive insights into membrane deformation.

%tube pullout BVP

%tube pullout BVP: Pinching

%%%%%%%%%%%%%%%%%%%%%%%%%%%%%%%%%%%%%%%%%%%%%%%%%%%
%%%%%%%%%%%%%%%%%%%%%%%%%%%%%%%%%%%%%%%%%%%%%%%%%%%
\subsection*{Piezo1-induced membrane footprint and gating response}
\begin{figure}[!htb]
\centering
  \psfrag{a}{(a) $\gamma = 1.0 \text{pN}/ \text{nm}$}
  \psfrag{b}{(b) $\gamma = 0.1 \text{pN}/ \text{nm}$}
  \psfrag{c}{(c) $\gamma = 0.05 \text{pN}/ \text{nm}$}
  \psfrag{d}{(d) $\gamma = 0.01 \text{pN}/ \text{nm}$}
  \psfrag{v}{A}
  \psfrag{w}{\small \shortstack{membrane \\footprint}}
  \psfrag{i}{\small membrane footprint}
  \psfrag{u}{view-A}
  \psfrag{x}{$x$}
  \psfrag{y}{$y$}
  \psfrag{z}{$z$}
  \psfrag{k}{\small $|u_y|$ (nm)}
  \psfrag{g}[c][c]{\small \shortstack{Rigid Piezo \\dome}}
  \psfrag{h}{$\phi$}
  \psfrag{l}{\small $u_y$(nm)}
\includegraphics[width=\textwidth]{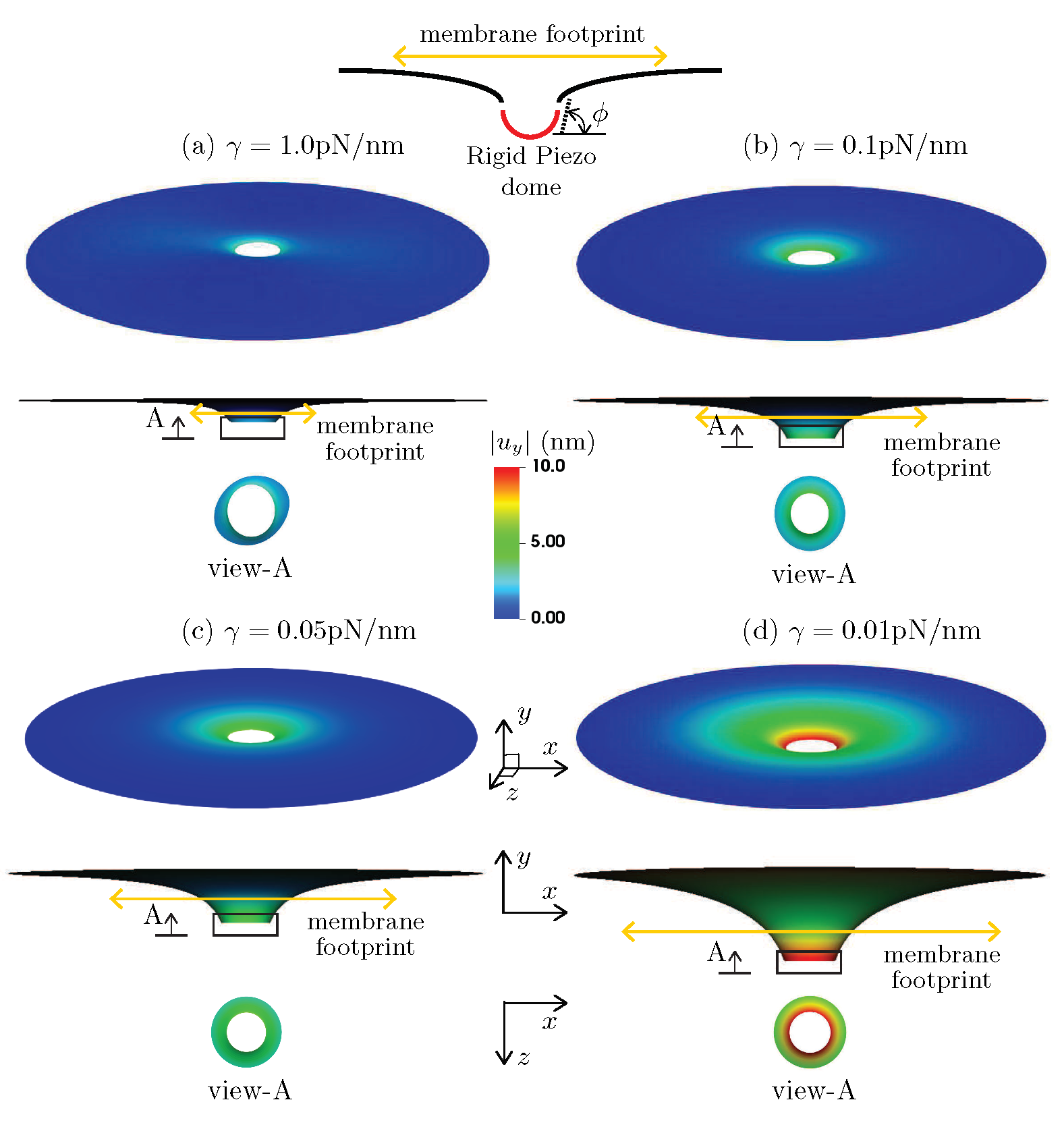}
\caption{Effect of surface tension on the membrane footprint area induced by a Peizo1 dome. Plotted are the 3D displacement profile, and its projection on the $x-y$ and $z-x$ planes. The bending modulus ($\kappa_B$) of the membrane is taken to be 30 pN$\cdot$nm, and a rigid Peizo dome effect is simulated by rotating the membrane (slope boundary condition) at the inner rim of the annular geometry to a value of $\phi=70$ degrees. To clearly visualize the increasing membrane footprint with decreasing surface tension, we scale the $y$ component of the displacement ($u_y$) by a factor of three in the $x-y$ oriented plots.}
\label{fig:peizoDomeBVP}
\end{figure}

We next investigate how mechanosensitive channels can deform the membrane.
Mechanosensitive ion channels on the cell membrane play an important role in the mechanosensory transduction processes of the cell. 
These ion channels are sensitive to the forces acting on the cell membrane and respond to these forces by undergoing conformational changes. 
These changes result in the opening and closing of pores in the cell membrane and thereby regulate the flow of ions and solutes entering and egressing the cell. 
Examples of such mechanosensitive ion channels include Piezo1, MscL and TREK-2 \cite{ridone2019piezo1}. 
In the case of Piezo1, a gated ion channel made up of three protein subunits that induce a dome-shaped structure on the cell membrane, the gating mechanism is triggered by the membrane surface tension. 
The membrane deformation induced by the surface tension acts as a mechanical signal that activates the protein subunits and causes them to undergo a conformational change that results in pore opening and transport of ions and solutes \cite{gottlieb2012gating,lewis2015mechanical,zhao2019mechanosensitive}.\\ 
While the exact mechanism of mechanosensory transduction effected by the Piezo1 ion channel is still an open question, the extent of the deformed shape induced by the Peizo1 dome (referred to as the membrane footprint) is understood to significantly influence the sensitivity of the gating response of the channel \cite{haselwandter2018piezo}. 
As observed by Haselwandter and MacKinnon \cite{haselwandter2018piezo}, an extended membrane footprint amplifies the sensitivity of Piezo1 subunits to respond to changes in the membrane surface tension. 
At the same time, increasing membrane tension significantly reduces the membrane footprint and thereby renders the Piezo1 subunits less sensitive to detect membrane mechanical signals. \\
In this analysis, we model the effect of surface tension on the area of the membrane footprint induced by the Piezo1 dome. 
The schematic for this boundary value problem is shown in Fig \ref{fig:schematict}(b), and the simulation results demonstrating the effect of surface tension on the membrane footprint are presented in \Cref{fig:peizoDomeBVP}. 
The plots show the 3D displacement profiles and their 2D projections under the boundary conditions enforced by the Piezo1 dome. 
A Piezo dome effect on the membrane is modeled by rotating the membrane (slope boundary condition) at the inner rim of the annular geometry to a value of $\phi=70$ degrees that is chosen so as to simulate the effect of a nearly hemispherical dome (which would correspond to $\phi=90$ degrees). 
This slope boundary condition assumes that the Piezo1 protein complex is a rigid dome that enforces a rotation on the surrounding membrane to ensure slope continuity between the hemispherical dome and the connected membrane. 
As can be seen from the subfigures \Cref{fig:peizoDomeBVP}(a)-(d), decreasing the surface tension increases the membrane footprint.
Especially in the limit of very low surface tension ($\gamma=0.01$ pN/nm) we see a significantly enhanced membrane footprint. 
The change in the out-of-plane displacement of the membrane, ($u_y$), shows a similar dependence on the surface tension. 
Since the out-of-plane displacement can be interpreted as a kinematic trigger to activate the gating mechanism in the protein subunits of Piezo1, this implies that at lower surface tension values, a higher value of $u_y$ is attained, thus delivering an amplified kinematic trigger, and therefore greater sensitivity of the Piezo1 dome to changes in surface tension. 
These results are consistent with the observations by Haselwandter and MacKinnon \cite{haselwandter2018piezo} that uses the classical reduced order Monge and arc-length axisymmetric parametrization methods to model the Piezo1-induced membrane deformation. Note that the deformation profile at the inner rim is, in general, non-axisymmetric, an effect that increases with membrane tension, $\gamma$. This illustrates the power of
 the 3D computational framework, which while it encompasses axisymmetric deformation, also admits non-axisymmetric modes. With access to the larger space, deformation profiles that are attainable at lower energies, and since the elasticity problem results in a (local-) minimum energy configuration, are indeed attained. Thus, while the 3D model reproduces the trends predicted by the reduced order models, its true power is in identifying more complex deformation patterns that are not accessible to the reduced order axisymmetric models.
%Using the 3D framework, we have noticed symmetry breaking in the membrane footprint shape resulting in a 3-fold radial symmetry membrane deformation profile instead of the axisymmetric deformation profile. However, we will explore this symmetric breaking behaviour in  detail in a future publication.         

%\todopr{Shiva, can we say something about how the triangular cross-section will likely induce a non-circular footprint in mechanical terms? Using the same language you've used so far?}

%%%%%%%%%%%%%%%%%%%%%%%%%%%%%%%%%%%%%%%%%%%%%%%%%%%
%%%%%%%%%%%%%%%%%%%%%%%%%%%%%%%%%%%%%%%%%%%%%%%%%%%
\subsection*{Budding of membranes by protein coats during endocytosis}
Budding of membranes by protein coats is a critical process in clathrin-mediated endocytosis (CME) that transports substance from the extracellular matrix to the cell interior.
Several key features, including protein-induced spontaneous curvature, membrane properties, membrane tension, and force from actin polymerization, have been identified to govern the bud formation in CME \cite{hassinger2017design,walani2015endocytic,nedelec2015}.
The ability to simulate the morphological progression of  bud formation in the 3D setting under different combinations of these identified features is crucial for understanding the mechanical progression of CME.
To further demonstrate the predictive capability of our simulation framework, we investigate the relationships between coat area, coat curvature, and degrees of symmetry during the budding of a vesicle as part of the endocytosis phenomena. 
The schematic of the simulation setup is given in Fig. \ref{fig:schematict}(c). 
We simulated  bud formation under two different conditions in our framework, similar to the setup explored in \cite{hassinger2017design}.
In the first case (i), the coated region has a fixed spontaneous curvature $h_0=0.02$ nm$^{-1}$ with progressively increasing area of the coat.
In the second case (ii), the coated region has a fixed radius of $80$ nm with progressively increasing spontaneous curvature. 
In both cases, the uncoated membrane has a radius of $400$ nm and has a surface tension with $\gamma=0.002$ pN/nm. The bending modulus ($k_B$) for all cases is taken to be $320$ pN$\cdot$nm. 
%Change Rotational constraints to slope boundary condition to be consistent with the earlier description. 
The slope boundary condition with $\phi=0$ is enforced at both the inner and outer rims of the membrane through the penalty method to ensure the continuous differentiability with the flat membrane reservoir. 
As illustrated in Fig. \ref{fig:schematict}(c),  Dirichlet boundary conditions are enforced to eliminate rigid body motions. 
The hyperbolic tangent function proposed in \cite{hassinger2017design, Rangamani2020} (see SI for illustration of this function) is used to ensure sharp but smooth transitions at the boundaries of the coated region and the uncoated membrane. 

\iffalse
\begin{table}[h!]
\begin{tabular}{llll}
  Parameter & Significance  & Value  & Ref(s) \\ \hline
  $\gamma$ & Edge membrane tension & 0.002 pN/nm   &  \cite{hassinger2017design}\\
  $\kappa$ & Bending rigidity of the membrane  & 320 pN$\cdot$nm & \cite{hassinger2017design,dimova2014recent} \\
  $\text{h}_0$ & Spontaneous curvature of coat & 0.02 nm$^{-1}$ & \cite{hassinger2017design,Miller2015,stachowiak2013cost} \\ \hline
\end{tabular}
\label{tab:parameters-spontaneous-curvature}
\caption{Parameters used in the bud formation simulation.}
\end{table}
%\todopr{This is different from what was used in the tube simulation. This value is closer to experimental measurements. comment on 0.002pN/nm in the table.}
\fi
%\todoxz{@All, should we still have tables to summarize parameters used in each example?}. SR: I think you already mentioned all the parameters that were in the table earlier in the text. Correct me if there are any parameters that are not mentioned in the text. 

\begin{figure}[!htb]
\centering
  \psfrag{X}[c][c]{\scriptsize $x$}
  \psfrag{Y}[c][c]{\scriptsize $y$}
  \psfrag{Z}[c][c]{\scriptsize $z$}
  \psfrag{view A}[c][c]{view A}
  \psfrag{view B}[c][c]{view B}
  \psfrag{H=0.0158}[c][c]{$h=0.0158$}
  \psfrag{1Area=22167}[l][c]{1: Area = 22167 (nm$^2$)}
  \psfrag{2Area=34636}[l][c]{2: Area = 34636 (nm$^2$)}
  \psfrag{3Area=44488}[l][c]{3: Area = 44488 (nm$^2$)}
  \psfrag{4Area=49876}[l][c]{4: Area = 49876 (nm$^2$)}
  \psfrag{5Area=54408}[l][c]{5: Area = 54408 (nm$^2$)}
  \psfrag{Hnm}[l][c]{$h$ (nm$^{-1}$)}
  \psfrag{H0}[c][c]{-0.002}
  \psfrag{H1}[c][c]{0.01}
  \psfrag{H2}[c][c]{0.02}
  \psfrag{Coat}[c][c]{Coat area (nm$^2$) $\times~10^4$ }
  \psfrag{Max}[c][c]{Max curv. (nm$^{-1}$)}
  \psfrag{Y1}[c][c]{0.020}
  \psfrag{Y2}[c][c]{0.015}
  \psfrag{Y3}[c][c]{0.010}
  \psfrag{Y4}[c][c]{0.005}
  \psfrag{Y5}[c][c]{0.000}
  \psfrag{1}[c][c]{1}
  \psfrag{2}[c][c]{2}
  \psfrag{3}[c][c]{3}
  \psfrag{4}[c][c]{4}
  \psfrag{5}[c][c]{5}
  \psfrag{X1}[c][c]{0}
  \psfrag{X2}[c][c]{2}
  \psfrag{X3}[c][c]{4}
\includegraphics[width=\textwidth]{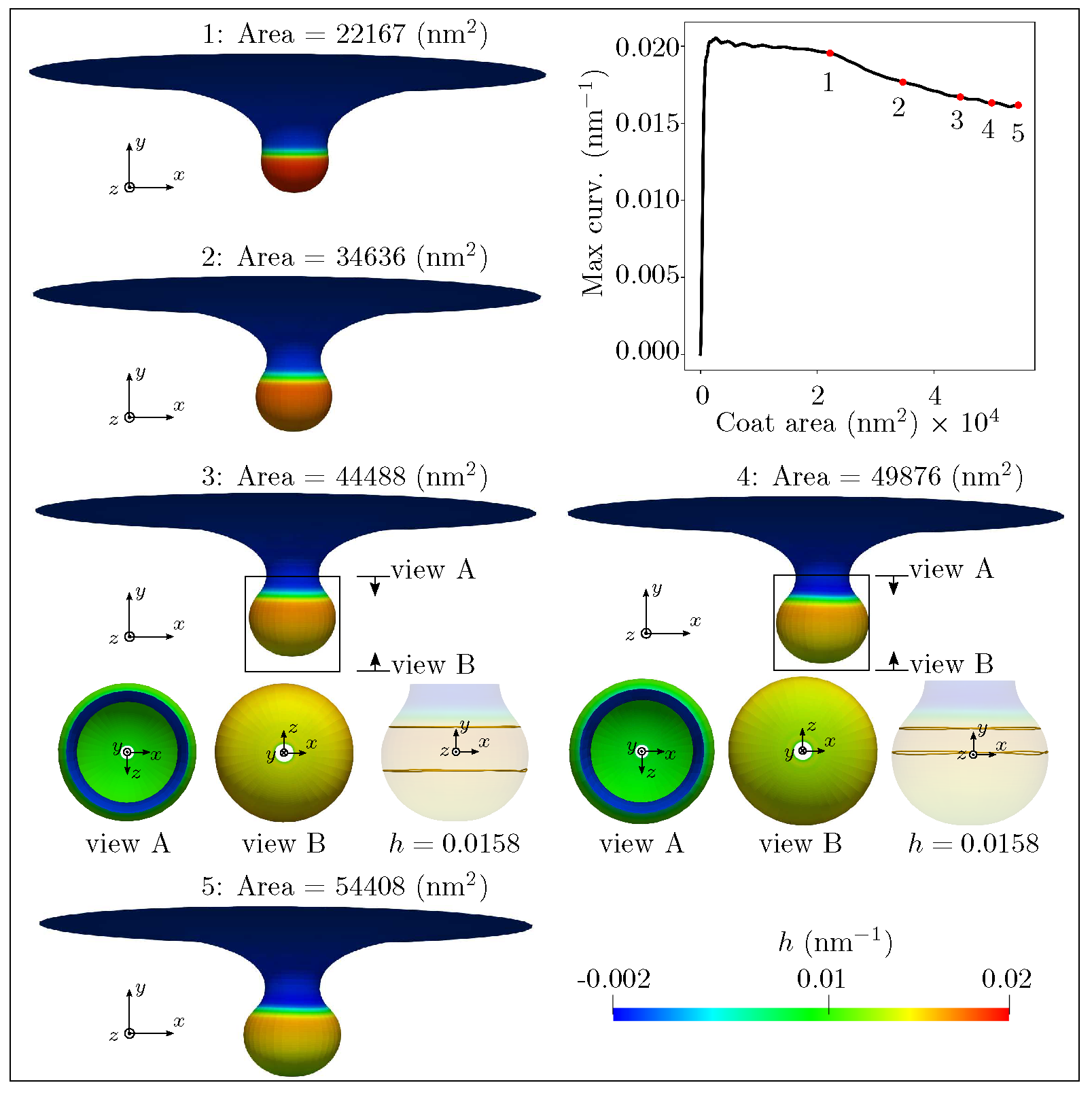}
\caption{
Formation of membrane buds with applied spontaneous curvature and increasing coat area.
A surface tension $\gamma=0.002$ pN/nm is applied at the outer rim of the membrane. The coated region has a spontaneous curvature $h_0=0.02$ nm$^{-1}$, corresponding to a curvature radius of $50$ nm. As illustrated by snapshots of the membrane at five different simulation stages each with increasing area, the membrane progresses from a flat shape to a bud-like shape with increasing coated area. The evolution of the maximum curvature curve is plotted, which is smooth throughout the simulation. Symmetry breaking is observed in this simulation, as the curvature contour plots at stage 3 and 4 with $h = 0.0158$ nm$^{-1}$ in the $x-y$ plane are not straight in the $y$ direction.}
\label{fig:spontaneous-A0}
\end{figure}

\begin{figure}[!htb]
\centering
  \psfrag{X}[c][c]{\scriptsize $x$}
  \psfrag{Y}[c][c]{\scriptsize $y$}
  \psfrag{Z}[c][c]{\scriptsize $z$}
  \psfrag{view A}[c][c]{view A}
  \psfrag{view B}[c][c]{view B}
  \psfrag{viewZ}[c][c]{zoom}
  \psfrag{H=0.018}[c][c]{$h=0.018$}
  \psfrag{1H=0.02800}[l][c]{1: $h_0$ = 0.028 (nm$^{-1}$)}
  \psfrag{2H=0.03220}[l][c]{2: $h_0$ = 0.0322 (nm$^{-1}$)}
  \psfrag{3H=0.03748}[l][c]{3: $h_0$ = 0.03748 (nm$^{-1}$)}
  \psfrag{4H=0.03756}[l][c]{4: $h_0$ = 0.03756 (nm$^{-1}$)}
  \psfrag{5H=0.04000}[l][c]{5: $h_0$ = 0.04 (nm$^{-1}$)}
  \psfrag{Hnm}[l][c]{$h$ (nm$^{-1}$)}
  \psfrag{H0}[c][c]{-0.003}
  \psfrag{H1}[c][c]{0.01}
  \psfrag{H2}[c][c]{0.02}
  \psfrag{H3}[c][c]{0.03}
  \psfrag{X0}[c][c]{0.00}
  \psfrag{X1}[c][c]{0.02}
  \psfrag{X2}[c][c]{0.04}
  \psfrag{Coat}[c][c]{Coat spontaneous curv. (nm$^{-1}$)}
  \psfrag{Max}[c][c]{Max curv. (nm$^{-1}$)}
  \psfrag{Y1}[c][c]{0.03}
  \psfrag{Y2}[c][c]{0.02}
  \psfrag{Y3}[c][c]{0.01}
  \psfrag{Y4}[c][c]{0.00}
  \psfrag{1}[c][c]{1}
  \psfrag{2}[c][c]{2}
  \psfrag{3}[c][c]{3}
  \psfrag{4}[c][c]{4}
  \psfrag{5}[c][c]{5}
\includegraphics[width=\textwidth]{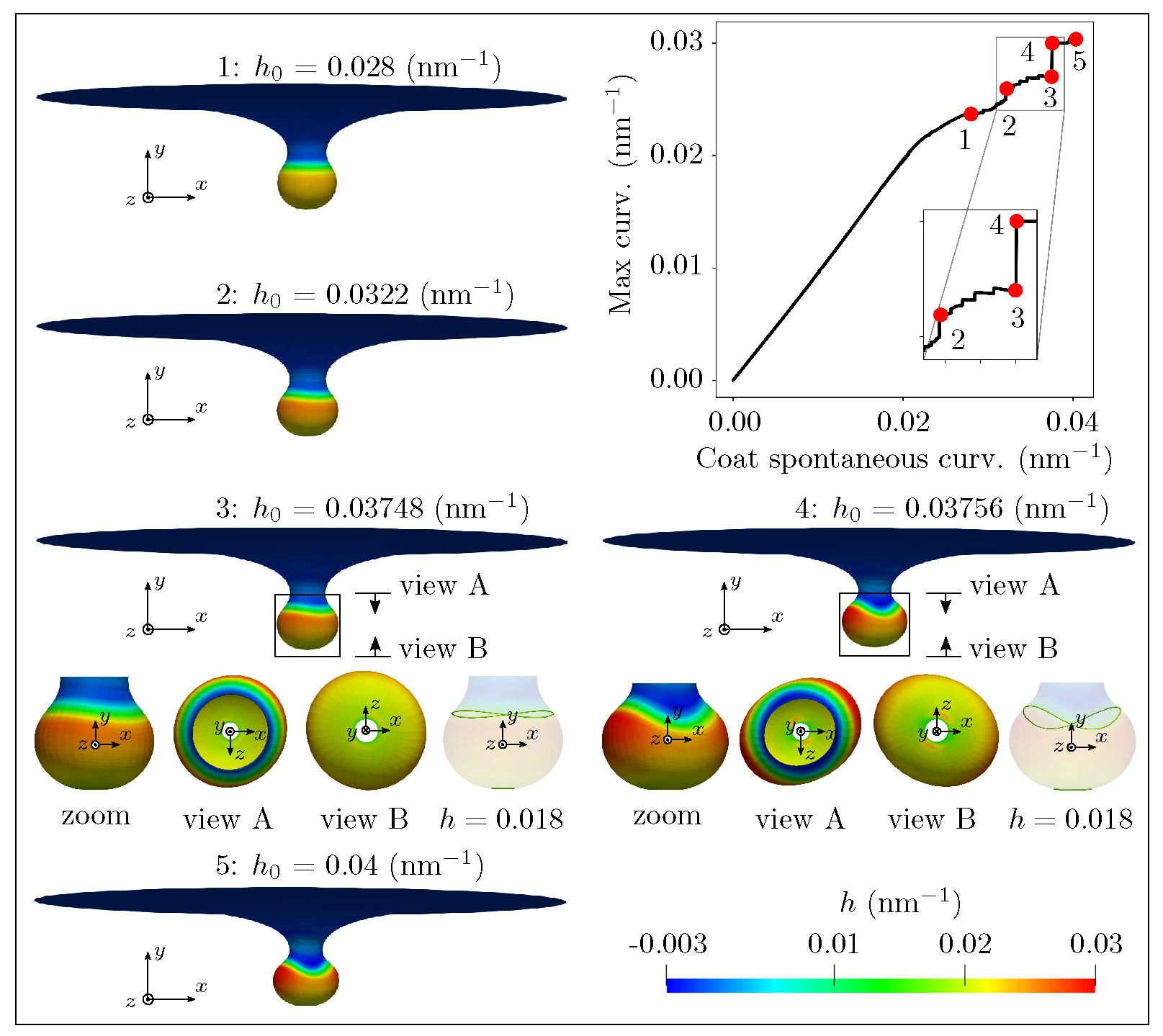}
\caption{ 
Formation of membrane buds with coat area and increasing spontaneous curvature.
A surface tension $\gamma=0.002$ pN/nm is applied  at the outer rim of the membrane. The coated region has a fixed radius of 80 nm. As illustrated by snapshots of the membrane at five different simulation stages each with increasing $h_0$, the membrane progresses from a flat shape to a bud-like shape with increasing $h_0$. The curved curvature contour plot in the $y$ direction in the $x-y$ plane at stage 3 with $h = 0.018$ nm$^{-1}$  indicates the existence of symmetry breaking in the simulation. The sudden change of the maximum curvature curve between stage 3 and 4 indicates a growth of the associated instability in this simulation setup, which is elaborated by the insets of simulation results at stage 3 and 4.}
\label{fig:spontaneous-H0}
\end{figure}

The simulation results from both setups are reported in Figs. \ref{fig:spontaneous-A0} and \ref{fig:spontaneous-H0}, where the coated membrane progresses from a flat shape to a bud-like shape. 
In addition, symmetry breaking of the membrane is observed in both cases. This is confirmed by the curved contour plot insets of $h$ in the $x-y$ plane in Figs. \ref{fig:spontaneous-A0} and \ref{fig:spontaneous-H0}, which would otherwise be straight lines, indicating constant heights in the $y$ direction. 
Figs. \ref{fig:spontaneous-A0} and \ref{fig:spontaneous-H0} further show that an additional instability mode exists in case (ii), where a rapid change in the maximum curvature curve between stage 3 and 4 occurs. 
It is worth mentioning that the spatial location of the maximum curvature evolves during the simulation. The different behavior of the maximum curvature curve in Figs. \ref{fig:spontaneous-A0} and \ref{fig:spontaneous-H0} indicate that the two simulations possess different energetic paths for their solutions.

\begin{figure}[!p]
\centering
{
  \psfrag{X-Z}[c][c]{$x-z$}
  \psfrag{X-Y}[c][c]{$x-y$}
  \psfrag{3D}[c][c]{3D}
  \psfrag{FHV}[c][c]{Fixed Coat $h_0$, Vary Coat Area}
  \psfrag{x0}[c][c]{\small 0}
  \psfrag{x1}[c][c]{\small 2}
  \psfrag{x2}[c][c]{\small 4}
  \psfrag{x3}[c][c]{\small 6}
  \psfrag{x4}[c][c]{\small 8}
  \psfrag{x5}[c][c]{\small 10}
  \psfrag{Coat}[c][c]{Coat area (nm$^{2}$) $\times$ $10^4$}
  \psfrag{Sp}[c][c]{Coat spontaneous curv. (nm$^{-1}$)}
  \psfrag{y0}[c][c]{\small 0.05}
  \psfrag{y1}[c][c]{\small 0.04}
  \psfrag{y2}[c][c]{\small 0.03}
  \psfrag{y3}[c][c]{\small 0.02}
  \psfrag{y4}[c][c]{\small 0.01}
  \psfrag{y5}[c][c]{\small 0.00}
  \psfrag{As}[l][l]{Asymmetry}
  \psfrag{ss}[l][l]{Symmetry}
\subfloat[
%each simulation has different but fixed H$_0$ with progressively increasing coated area
\label{fig:phase-diagram-vary-a}]{
\includegraphics[width=0.75\textwidth]{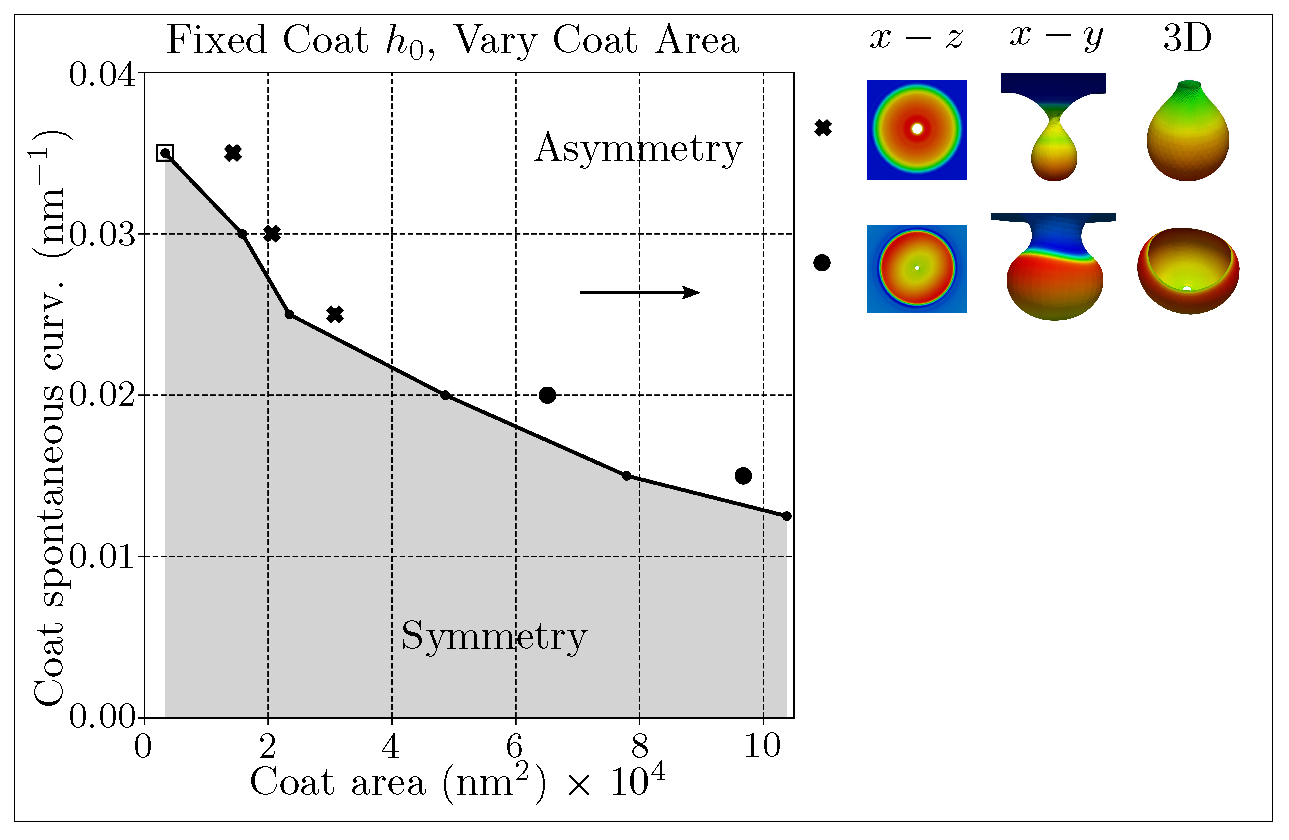}
}}\\
{
  \psfrag{X-Z}[c][c]{$x-z$}
  \psfrag{X-Y}[c][c]{$x-y$}
  \psfrag{3D}[c][c]{3D}
  \psfrag{FVH}[c][c]{Fixed Coat Area, Vary Coat $h_0$}
  \psfrag{x0}[c][c]{\small 0}
  \psfrag{x1}[c][c]{\small 2}
  \psfrag{x2}[c][c]{\small 4}
  \psfrag{x3}[c][c]{\small 6}
  \psfrag{x4}[c][c]{\small 8}
  \psfrag{x5}[c][c]{\small 10}
  \psfrag{Coat}[c][c]{Coat area (nm$^{2}$) $\times$ $10^4$}
  \psfrag{Sp}[c][c]{Coat spontaneous curv. (nm$^{-1}$)}
  \psfrag{y0}[c][c]{\small 0.05}
  \psfrag{y1}[c][c]{\small 0.04}
  \psfrag{y2}[c][c]{\small 0.03}
  \psfrag{y3}[c][c]{\small 0.02}
  \psfrag{y4}[c][c]{\small 0.01}
  \psfrag{y5}[c][c]{\small 0.00}
  \psfrag{As}[l][l]{Asymmetry}
  \psfrag{ss}[l][l]{Symmetry}
\subfloat[
%each simulation has different but fixed coated area with progressively increasing H$_0$
\label{fig:phase-diagram-vary-h}]{
\includegraphics[width=0.75\textwidth]{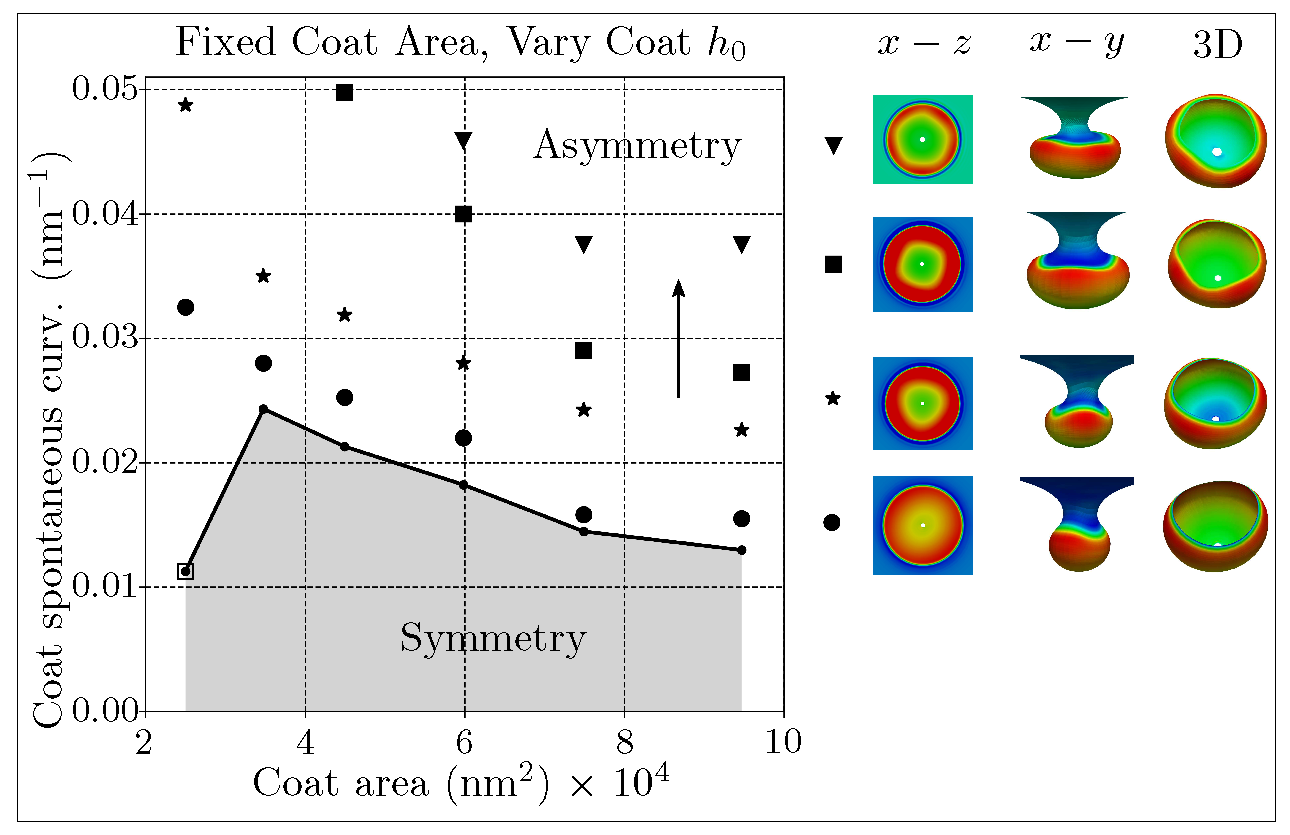}
}}
\caption{Membrane morphology evolution phase diagram for (a) similar simulation setup as in Fig. \ref{fig:spontaneous-A0} with fixed discrete $h_0$ but increasing protein-coated area, (b) similar simulation setup as in Fig. \ref{fig:spontaneous-H0} with fixed discrete protein-coated area but increasing $h_0$. 
%\todokg{Do you mean the direction in which the modes appear? Not clear what quantities.} 
The arrows indicate the progressively increasing quantities. The asymmetry morphology patterns differ for these two simulation setups.  For case (a), both twisting (cross) and two-fold (dot) wave shapes were captured in the asymmetry region. For case (b), two-fold (dot), three-fold (star), four-fold (square), and five fold (triangle) shapes were captured in the asymmetry region. Representative bud shapes colored by curvature values are shown in different views, where $x-z$ view is in the undeformed configuration, $x-y$ view and 3D view are in the deformed configuration. The dots with empty surrounding square indicate the cases where the proposed symmetry breaking criterion does not hold. 
%\XZ{@Shiva, I used $X-Z$ instead of $x-z$ because the first column view is in the undeformed configuration.}
}
\label{fig:spontaneous-phase-diagram}
\end{figure}

Next, we conducted six simulations for each case to construct the membrane morphology evolution phase diagram. In each simulation of case (i), a different value of $h_0$ is assigned to the coated region. 
For a given value of $h_0$ we progressively increase the area of the coated region at an identical increment for all the simulations to allow the bud to form. 
In each simulation of case (ii), the radius of the coated region was set to a different value and then $h_0$ of the coated region was progressively increased at an identical increment for all the simulations to allow the bud to form. 
The membrane morphology evolution phase diagram for both simulation setups appears in Fig. \ref{fig:spontaneous-phase-diagram} with arrows indicating progressively increasing quantities, where different patterns appear in the asymmetric region. 
To detect the symmetry breaking in each simulation, first, we uniformly sample 20 discrete values of $u_y$ between its minimum and maximum at each increment of the coat area for case (i) or the coat $h_0$ for case (ii). 
Next, the range of the curvature $h$, $[h_\text{min}, h_\text{max}]$,  at every discrete $u_y$ is computed. 
For those heights with $h_\text{min} > 0$, the relative change of $h$,  denoted as $\Delta h = 2(h_\text{max} -h_\text{min} )/(\text{abs}(h_\text{max}) +\text{abs}(h_\text{min}))$, is computed. 
At each incremental step, we thus have multiple values of $\Delta h$.  
Then the median value of $\Delta h$, denoted as $\Delta h_\text{med}$, is computed for that step. 
Now, for each simulation, we have an array of $\Delta h_\text{med}$, whose length is equal to the total number of incremental steps of either the coat area or the coat $h_0$. 
Our results show that symmetry-breaking usually can be detected when $\Delta h_\text{med}$ is at its minimum over increments of coat area or $h_0$, pointing to a close to uniform value of $\Delta h$ for that increment. \footnote{We remark that this is a loose criterion for detecting symmetry breaking, as not all the symmetry-breaking events occur precisely at the loading step where $\Delta h_\text{med}$ is minimal. 
However, this criterion generally provides a consistent indication of symmetry-breaking compared with our visual observation. 
Detailed discussion on the chosen symmetry-breaking criteria is provided in the SI.}. 
The associated values of $h_0$ and the area of the coated region at the specific incremental step where $\Delta h_\text{med}$ is at its minimum are used to construct the phase diagram in Fig. \ref{fig:spontaneous-phase-diagram}. 
The dots with an empty surrounding square in Fig. \ref{fig:spontaneous-phase-diagram} indicate the cases where the proposed symmetry breaking criterion does not hold. 
In the asymmetry region, not all the asymmetric patterns could be captured by the numerical simulations due to loss of numerical convergence during the solution iterations. 
This is due to the ill-conditioning of the system Jacobian matrix, that in turn is caused by the severe geometric and material non-linearity, potentially including bifurcation points, in the vicinity of asymmetric deformation modes. 
We use standard unconstrained optimization methods like arc-length and trust-region to achieve convergence of the solution iterations, whenever possible, and report successfully captured asymmetric patterns. 
The placement of the reported patterns are shown in Fig. \ref{fig:spontaneous-phase-diagram}. 
Here, the same markers are chosen to denote similar shapes. The fact that our computational framework could capture the symmetry breaking behavior, and even the pattern changes from 2-fold to 3-fold/4-fold/5-fold, demonstrates the advantages of the proposed 3D model over a reduced order axisymmetric model \cite{hassinger2017design, walani2015endocytic,alimohamadi2018role}.

%%%%%%%%%%%%%%%%%%%%%%%%%%%%%%%%%%%%%%
%%%%%%%%%%%%%%%%%%%%%%%%%%%%%%%%%%%%%%
%%%%%%%%%%%%%%%%%%%%%%%%%%%%%%%%%%%%%%
%%%%%%%%%%%%%%%%%%%%%%%%%%%%%%%%%%%%%%
%%%%%%%%%%%%%%%%%%%%%%%%%%%%%%%%%%%%%%
%%%%%%%%%%%%%%%%%%%%%%%%%%%%%%%%%%%%%%
%%%%%%%%%%%%%%%%%%%%%%%%%%%%%%%%%
\section*{Conclusion}
\label{sec:discussion}
Biomembranes play a central role in various cell-scale and organelle-scale phenomena like locomotion of cells \cite{zhao2013exo70}, packaging and trafficking of nutrients and signalling constituents \cite{liu2009mechanochemistry}, maintaining organelle morphology and functionality \cite{hu2008membrane, shibata2006rough, mcniven2006vesicle}, etc. 
In almost all these processes, these surfaces are known to undergo significant deformation through bending; and the evolution of the out-of-plane bending deformation is a key mechanism of morphological evolution, besides in-plane fluidity. Thus, many analytical and numerical approaches exist in the literature to model bending and curvature generation, especially for solving the governing equations resulting from the Helfrich-Canham \cite{helfrich1973elastic} characterization of membrane elasticity. \\
While these widely used analytical and numerical approaches (e.g. Monge parametrization, arclength parametrization and asymptotic methods) yield solutions to a wide range of boundary value problems of membrane bending, they are intrinsically limiting in capturing the complete envelope of membrane deformations due to the underlying axisymmetric restrictions on the kinematics and boundary conditions. 
Since the study of biomembrane deformation draws heavily from the well established models of elastic shells \cite{helfrich1973elastic, novozilov1959}, it is only natural to look for the validity of axisymmetric approximations and for the existence of non-axisymmetric solutions in the deformation of elastic shell geometries. 
Interestingly, many elastic structures have inherent modes of instability that result in enhanced deformation or even collapse in response to loading and are associated with lower energy barriers.
Such modes are ubiquitous in thin elastic shells and manifest as folding, wrinkling, creasing, and buckling (e.g. wrinkling of thin membranes and graphene sheets \cite{Deng2016}, surface tension induced buckling of liquid-lined elastic tubes \cite{hazel2005surface}, snap-through of elastic columns \cite{brojan2007buckling}, barrelling modes of thin cylinders \cite{Azzuni2018, Rahman2011}, etc.). 
Notably, they have lower symmetry than the fully axisymmetric deformations. 
If such modes exist, and are accessible in biomembranes, then they would naturally lead to a reduction in the load and energy barriers to membrane deformation, and may result in heretofore numerically unexplored deformation profiles and membrane morphologies. 
Accessing these lower symmetry modes and predicting the complex, three dimensional deformation profiles in biomembranes provided the primary motivation for developing the computational framework presented in this work. \\
We note that
the first application of this framework was in our recent study demonstrating how non-axisymmetric buckling lowers the energy barrier associated with membrane neck constriction in biomembranes \cite{Vasan2019}. 
In that study, we used a mechanical model of the lipid bilayer to systematically investigate the influence of location, symmetry constraints, and helical forces on membrane neck constriction. Simulations from our model demonstrated that the energy barriers associated with constriction of a membrane neck are location-dependent and are significantly affected by kinematic constraints on the deformation. 
Importantly, if symmetry restrictions on the membrane deformation are relaxed, the constriction pressure and thus the energy barrier for constriction are dramatically lowered. 
Our studies established that despite different molecular mechanisms of neck formation in cells, the mechanics of constriction naturally leads to a loss of symmetry and occurs at a much lower load/energy threshold. 
Motivated by the improved understanding of membrane deformation and the undesired effects of axisymmetry restrictions observed in that study, we have further developed the framework and expanded its scope to modeling other important membrane deformation processes.\\

Accordingly, in this work, we model three classical biomembrane problems: formation of tubular shapes and their lateral constriction, Piezo1-induced membrane footprint generation, and budding of membranes during endocytosis. 
For each of these problems, we are able to validate against results and observation available in the literature for the simpler deformation modes, and also predict the more complex, less symmetric deformation profiles that are not accessible by the traditional analytical methods and axisymmetric numerical methods. 
Moreover, for the problem of endocytic vesicle budding, we also map a phase diagram classifying the symmetric and less-symmetric states.\\  

The computational framework is implemented as an open-source software library and provided as a resource to the biophysics community. It is expected that this framework will serve as a platform for exploring complex deformation mechanisms (including geometric bifurcations and post-bifurcation responses) in biomembranes, and result in an improved understanding of the mechanics underlying various biomembrane phenomena. 
Future extensions envisioned are support for in-plane fluidity \cite{rangamani2013interaction}, surface diffusion (to model protein transport on the membrane) 
%\cite{Mahapatra2020}
, and a contact model (to model membrane-membrane interactions).
In addition, the inability of the current framework to apply non-uniform Dirichlet boundary conditions and constraints on displacement degrees of freedom inside the domain (i.e. at non-interpolatory knots of the spline surface) are significant limitations and will be addressed in future developments.    

%\afterpage{\clearpage}
\section*{Acknowledgments}
D. A., R. G. and S. R. would like to acknowledge the Wisconsin Alumni Research Foundation (WARF) and the Grainger Institute for Engineering at UW-Madison for funding support.\\
X. Z. and K. G. would like to acknowledge the support of computing resources provided in part by the National Science Foundation, United States via grant 1531752 MRI: Acquisition of Conflux, A Novel Platform for Data-Driven Computational Physics (Tech. Monitor: Ed Walker). \\
P. R. was supported by NIH R01GM132106 for this work. 

%\section*{References}
\bibliography{ref}
%

%%%%%%%%%%
%%%%%%%%%%
%%%%%%%%%%
\end{document}